\documentclass[12pt,authoryear]{elsarticle}

\usepackage{hyperref}
\usepackage{graphicx}
\usepackage{subcaption}
\usepackage{amssymb}
\usepackage{amsmath}
\usepackage{multirow}
\usepackage[utf8]{inputenc}
\usepackage[depythontex]{pythontex}
\usepackage{pgfplots}
\usepackage{pgfplotstable}
\usepgfplotslibrary{groupplots}
\usepackage{float} 
\usepackage{derivative}
\usepackage{latexcolors} 
\usepackage{import} 
\usepackage{siunitx}
\usepackage{catchfile}
\usepackage{mathrsfs} 
\usepackage{standalone} 
\usepackage{makecell}
\usepackage{algorithm}
\usepackage{algpseudocodex}
\usepackage{cleveref}
\usepackage{lineno}
\usepackage{varwidth}

\pgfplotsset{compat=newest, 
	table/search path={figs} 
}

\usepgfplotslibrary{external}
\usetikzlibrary{math, shapes.geometric, shapes.misc, calc, arrows.meta, fit} 
\usepackage{tcolorbox}

\tikzsetexternalprefix{figcache/}
\tikzset{external/mode=list and make}

\usepackage[abbreviations]{glossaries-extra}
\setabbreviationstyle{long-short} 

\usepackage{xcolor}
\usepackage{xargs}
\usepackage[colorinlistoftodos,textsize=footnotesize]{todonotes}
\usepackage{lineno}

\newcommandx{\unsure}[2][1=]{\todo[linecolor=red,backgroundcolor=red!25,bordercolor=red,#1]{#2}}
\newcommandx{\change}[2][1=]{\todo[linecolor=blue,backgroundcolor=blue!25,bordercolor=blue,#1]{#2}}
\newcommandx{\info}[2][1=]{\todo[linecolor=OliveGreen,backgroundcolor=OliveGreen!25,bordercolor=OliveGreen,#1]{#2}}

\usepackage{array}
\newcommand{\PreserveBackslash}[1]{\let\temp=\\#1\let\\=\temp}
\newcolumntype{C}[1]{>{\PreserveBackslash\centering}p{#1}}
\newcolumntype{R}[1]{>{\PreserveBackslash\raggedleft}p{#1}}
\newcolumntype{L}[1]{>{\PreserveBackslash\raggedright}p{#1}}

\tikzstyle{startstop} = [rectangle, rounded corners, 
minimum width=5cm, 
minimum height=1cm,
text centered, 
draw=black, 
fill=red!30]

\tikzstyle{io} = [trapezium, 
trapezium stretches=true, 
trapezium left angle=70, 
trapezium right angle=110, 
minimum width=5cm, text width=5cm, 
minimum height=1cm, text centered, 
draw=black, fill=blue!30]

\tikzstyle{process} = [rectangle, 
minimum width=5cm, 
minimum height=1cm, 
text centered, 
text width=5cm, 
draw=black, 
fill=orange!30]

\tikzstyle{decision} = [diamond, 
minimum width=3cm, 
minimum height=0.75cm, 
text centered, 
draw=black, 
fill=green!30]
\tikzstyle{arrow} = [thick,->,>=stealth]

\tikzstyle{pprocess} = [
    draw,
    minimum width=3.5cm,
    minimum height=1.25cm,
    outer sep=0,
    rectangle,
    rounded corners,
    append after command={
        \pgfextra{
            \draw (\tikzlastnode.north west) -- ++ (-0.3,0) |- (\tikzlastnode.south west);
            \draw (\tikzlastnode.north east) -- ++ (0.3,0) |- (\tikzlastnode.south east);
        }
    }]

\newabbreviation
{sph}
{SPH}
{Smoothed Particle Hydrodynamics}

\newabbreviation
{wcsph}
{WCSPH}
{Weakly Compressible \gls{sph}}

\newabbreviation
{isph}
{ISPH}
{Incompressible \gls{sph}}

\newabbreviation
{mms}
{MMS}
{Method of Manufactured Solutions}

\newabbreviation
{sisph}
{SISPH}
{Simple Iterative \gls{sph}}

\newabbreviation
{dtsph}
{DTSPH}
{Dual-Time \gls{sph}}

\newabbreviation
{edac}
{EDAC}
{Entropically Damped Artificial Compressibility}

\newabbreviation
{dbc}
{DBC}
{Dynamic Boundary Condition}

\newabbreviation
{mdbc}
{mDBC}
{modified \gls{dbc}}

\newabbreviation
{lust}
{LUST}
{Local Uniform STencil}

\newabbreviation
{tvdrk2}
{TVD-RK2}
{Total Variaton Diminishing - Runge Kutta 2}

\newabbreviation
{fvm}
{FVM}
{Finite Volume Method}

\newabbreviation
{fou}
{FOU}
{First Order Upwind}

\newabbreviation
{mls}
{MLS}
{Moving Least Squares}

\newabbreviation
{tvf}%
{TVF}%
{Transport Velocity Formulation}

\newabbreviation
{epec}%
{EPEC}%
{Evaluate Predict Evaluate Correct}

\newabbreviation
{mi1}
{MI1}
{Matrix Inversion formulation 1}

\newabbreviation
{cfd}
{CFD}
{Computational Fluid Dynamics}

\newabbreviation
{eno}
{ENO}
{Essentially Non-Oscillatory}

\newabbreviation
{weno}
{WENO}
{Weighted \gls{eno}}

\newabbreviation
{muscl}
{MUSCL}
{Monotone Upstream-Centered Schemes for Conservation Laws}
\journal{}

\begin{document}

\begin{frontmatter}

	\title{Adaptive Compressible Smoothed Particle Hydrodynamics}
	\author[IITB]{Navaneet Villodi\corref{cor1}}
	\ead{navaneet@iitb.ac.in}
	\author[IITB]{Prabhu Ramachandran}
	\ead{prabhu@aero.iitb.ac.in}
	\address[IITB]{Department of Aerospace Engineering, Indian Institute of
		Technology Bombay, Powai, Mumbai 400076}

	\cortext[cor1]{Corresponding author}

	\begin{abstract}
		Modulating the number of particles in a region is key to accurately capturing the nuances in compressible flows with \gls{sph}.
		This paper presents a volume-based adaptive refinement and derefinement procedure, with state-of-the-art features such as automatic local adaptivity and solution adaptivity applied in the context of compressible flows.
		A shock-aware particle shifting procedure is introduced to regularize the particle distribution while preserving the integrity of shocks.
		To our knowledge, this is the first demonstration of shock-based solution adaptivity and shock-aware particle shifting in the literature.
		A wide variety of test problems, which involve flow in and around boundaries, are employed to highlight the utility of these adaptivity features in improving the results and in making simulations faster.
		For instance, the adaptive resolution procedure is shown to achieve an order of magnitude increase in computational speed.
		We also demonstrate the effectiveness of the adaptivity procedure in resolving issues such as errors arising from the interaction with differently spaced ghost particles at boundaries, formation of spot-like structures due to particle clumping, and poorly resolved low-density regions.
		In essence, the adaptivity technique presented in this paper is a powerful tool for simulating compressible flows with enhanced accuracy and efficiency.
	\end{abstract}


\end{frontmatter}
\section{Introduction} \label{sec:intro}

\gls{sph} is a numerical method that can be used to simulate various physical phenomena.
While convergence, consistency and stability, boundary treatment are considered as challenges in \gls{sph}~\citep{vacondioGrandChallengesSmoothed2021}, many of its advantages, like the Lagrangian nature, ease of dealing with moving bodies, and ability to handle free surfaces make it an attractive choice for many engineering applications involving compressible flows.

Compressible hydrodynamics has been one of the focal points of \gls{sph} since its origins~\citep{monaghanShockSimulationParticle1983}. Artificial viscosity and artificial conductivity have been used for stabilization and shock capturing. Various switches have been introduced to limit excess dissipation away from shocks~\citep{balsaraNeumannStabilityAnalysis1995, morrisModelingLowReynolds1997, cullenInviscidSmoothedParticle2010,readSPHSSmoothedParticle2012,frontiereCRKSPHConservativeReproducing2017, rosswogSimpleEntropybasedDissipation2020}. While artificial viscosity can be regarded as an approximate Reimann solver~\citep{monaghanSPHRiemannSolvers1997, puriApproximateRiemannSolvers2014, priceShockCapturingSmoothed2024}, Godunov variants of \gls{sph} have also been worked upon in the context of compressible flows~\citep{inutsukaReformulationSmoothedParticle2002,
chaKelvinHelmholtzInstabilitiesGodunov2010,
iwasakiSmoothedParticleMagnetohydrodynamics2011,muranteHydrodynamicSimulationsGodunov2011,
puriApproximateRiemannSolvers2014}. A comparison by \citet{puriComparisonSPHSchemes2014} showed that the results of an artificial viscosity based formulation~\citep{priceSmoothedParticleMagnetohydrodynamics2004} is comparable to Godunov \gls{sph}. 

More recently, there has been a considerable focus on high order \gls{sph} schemes with \gls{muscl} and \gls{eno} based approaches~\citep{koukouvinisImprovedMUSCLTreatment2013, avesaniNewClassMovingLeastSquares2014,nogueiraHighaccurateSPHMethod2016,avesaniAlternativeSPHFormulation2021,pengEfficientTargetedENO2021,wangNewTypeWENO2021, mengTargetedEssentiallyNonoscillatory2022, gaoNewSmoothedParticle2023,antonaWENOSPHScheme2023}. While these schemes can achieve sharper resolution of shocks, they are computationally still limited to first order accuracy near discontinuities for enforcing stability and monotonicity, according to Godunov's theorem~\citep{toroRiemannSolversNumerical2009}. Adaptive refinement, therefore, emerges as a powerful tool for sharper resolution of shocks and other features in compressible flows.

\gls{isph} and \gls{wcsph} have found adoption in engineering applications~\citep{shadlooSmoothedParticleHydrodynamics2016}.
However, applications of compressible \gls{sph} have been mostly restricted to astrophysics~\citep{rosswogAstrophysicalSmoothParticle2009,springelSmoothedParticleHydrodynamics2010a} and magnetohydrodynamics~\citep{priceSmoothedParticleHydrodynamics2012}.
We conjectured that the lack of adoption is due to deficiency of robust solid boundary treatment methods for compressible \gls{sph}.
On that account, we proposed boundary treatment methods for compressible \gls{sph}~\citep{villodiRobustSolidBoundary2024} by modifying existing boundary treatment methods widely used with \gls{wcsph} and \gls{isph}.
A significant conclusion of this work was also that modulating the number of particles in a region is not just a convenience but a necessity in certain compressible flow scenarios. Listed below are some scenarios which can be effectively resolved by incorporating an adaptive resolution procedure:

\begin{enumerate}
	\item Resolution deficiency: The wake formed behind a blunt body in compressible flows, or an expansion in general can lead to formation of a low density region.
	      As particle density is tied to fluid density, the features in the low-density region would be poorly resolved. \citet{vacondioShallowWaterSPH2013} employed particle splitting and merging to approach a similar problem with shallow water flows. However, this has never been addressed satisfactorily in compressible \gls{sph} literature, to our knowledge.
	\item Ghost-particle interaction: Ghost particles representing bodies are set up with a fixed spacing or representative volume.
	As the flow evolves, fluid particles near the boundary could end up with a very different volume.
	      This inconsistency could lead to errors in the interaction between fluid and ghost particles, increasing the propensity of fluid particles to leak through the ghost particles.
	      This represents a significant challenge in compressible flows but is rarely addressed in the literature. See \citet{villodiRobustSolidBoundary2024} for a discussion on this.
	\item Particle clumping: Maintaining appropriate particle volume is critical for mitigating errors arising from particle inconsistency. As the particle density can change significantly across a shock wave, particles with very different associated volumes could end up interacting with each other within the fluid itself.
	      This could lead to clumping of particles and formation of spot-like structures. This is a well-studied issue in SPH literature~\citep{swegleSmoothedParticleHydrodynamics1995,dehnenImprovingConvergenceSmoothed2012}.
\end{enumerate}
While various techniques have been proposed to particle clumping~\citep{monaghanSPHTensileInstability2000,dehnenImprovingConvergenceSmoothed2012,lahiriStableSPHAdaptive2020}, we show that the adaptive resolution procedure presented in this work is also effective at remedying this issue. We also demonstrate that the other two issues described above can be effectively resolved by the adaptive resolution procedure. The algorithm builds upon existing state-of-the-art techniques in literature. Our adaptive resolution procedure for highly-compressible supersonic and hypersonic flows is the first of is kind, as evidenced by quality results over a wide over a wide range of problems, most of which are not found in Lagrangian compressible \gls{sph} literature. The following are some novel features that could be regarded as the highlights of this work:

\begin{enumerate}
	\item Shock-based solution adaptivity: Particles near shocks are identified and are then automatically refined for capturing sharper shocks.
	      This is a novel feature of the present work and is being demonstrated in literature for the first time, to our knowledge.
	\item Scheme independent adaptivity procedure: The adaptivity procedure is demonstrated with two different discretization schemes to show that it can be applied atop any existing discretization scheme with minimal or no modifications.
	\item Shock-aware shifting: A procedure is introduced to ensure that particles near shocks are not disturbed during the particle shifting process.
	This, to our knowledge, is being demonstrated in the literature for the first time.
    \item Comprehensive test cases: A variety of test problems are used to demonstrate the effectiveness of the adaptive resolution procedure in remedying existing issues and producing better and faster results. Except for the oscillating piston chamber problem, which serves as a verification test case, none of the presented problems have been previously solved using \gls{sph} with adaptive resolution by other researchers, to the best of our knowledge.
	\item Source code availability: The source code for the adaptivity procedure and rest of this work is made available in the interest of reproducibility and to encourage further research in this direction at \url{https://gitlab.com/pypr/adaptive-compressible-sph}.
\end{enumerate}

The various features of the adaptive resolution procedure can be applied independently or in conjunction with each other to suit the needs of the problem at hand. Most importantly, the adaptive resolution allows us to obtain better results with fewer particles.  Therefore, by leveraging adaptive resolution, we can significantly accelerate computations while minimizing the use of computational resources. For instance, in the rotating square projectile problem~(\cref{sec:rotsq}), the adaptive resolution procedure delivered a ten fold speedup in the simulation time.
This potential for acceleration may be even greater for more complex practical problems where the intricate details of the flow field needs to be captured in a localized region, while the rest of the domain can be kept coarser.  

The rest of the paper is organized as follows.
In \cref{sec:lit}, we provide a brief review of the relevant literature on adaptive resolution in \gls{sph}.
The underlying discretization scheme upon which the proposed adaptivity procedure is implemented is described in \cref{sec:formulation}.
In \cref{sec:method}, we introduce the proposed adaptivity procedure and provide a detailed explanation of its implementation.
The procedure is applied to a variety of test problems, and the results are presented in \cref{sec:results}.
Finally, we summarize and conclude the work with \cref{sec:conclusions}.
 
\section{State of adaptive resolution in \gls{sph}} \label{sec:lit}
Early ideas on adaptive refinement in \gls{sph} are found in the works of \citet{kitsionasSmoothedParticleHydrodynamics2002} and \citet{lastiwkaAdaptiveParticleDistribution2005}.
\citet{kitsionasSmoothedParticleHydrodynamics2002} consider replacing a parent particle with a number of offspring particles.
Merging of particles is not considered.
On the other hand, \citet{lastiwkaAdaptiveParticleDistribution2005} consider introducing new particles into the domain without removing the parent particle, adjusting the positions iteratively and assigning properties by interpolating them from existing particles.
Particles to be removed are deleted.
The mass and volume of the particles are reallocated upon splitting and removal.

The work of \citet{feldmanDynamicRefinementBoundary2007} should be regarded as seminal as they clearly laid down constraints on conservation and the strategy to minimize error in density profiles while splitting of particles.
They did not consider derefinement of particles.
\citet{vacondioSmoothedParticleHydrodynamics2012} extended \citeauthor{feldmanDynamicRefinementBoundary2007}'s work and also augmented their procedure with a derefinement technique~\citep{vacondioVariableResolutionSPH2013}.

The work of \citet{omidvarWaveBodyInteraction2012} investigates the feasibility of having different refinement regions.
The region of interest, say the region around a body, is refined by having more number of low-mass particles.
However, in this work, the splitting and merging of particles is not dynamic; the predefined refinement region is just initialized as such.
This work identified particle clumping at interface of refinement regions, where particles of significantly different masses interact.
The work of \citet{khorasanizadeDynamicFlowbasedParticle2016} is another instance where significant clumping artefacts were observed at the interface of refinement regions.
\citet{omidvarWaveBodyInteraction2012} had already reported that such clumping can be avoided by achieving the desired resolution in multiple refinement steps, for example, not letting particles with mass \(m\) directly interact with particles of mass \(m/16\) by introducing an intermediate refinement layer with particles of mass \(m/4\).

Later, \citet{yangSmoothedParticleHydrodynamics2017} introduced a technique to automatically define refinement layers for a smoother transition between regions of different resolutions.
They also introduced a new splitting procedure in which a particle is split into two, unlike \citet{vacondioVariableResolutionSPH2013} where a particle is split into seven.
They demonstrated the application of adaptive resolution in multiphase flows~\citep{yangAdaptiveResolutionMultiphase2019}.
\citet{liEfficientNoniterativeSmoothed2023} reports that the procedure introduced by \citet{yangAdaptiveResolutionMultiphase2019} to adaptively set the smoothing length exhibits poor stability.

\citet{barcaroloAdaptiveParticleRefinement2014} introduced a novel approach involving the retention of the parent particle after creating offspring particles.
As the parent particle enters a refinement region, offspring particles are created.
The parent particle is not deleted but deactivated; it does not interact with other particles.
It is advected with the time derivatives computed using interpolation based on the offspring particles.
As the parent particle exits the refinement region, it is reactivated and the offspring particles are erased.
Later, \citet{sunSPHModelSimple2017} added this procedure to their \(\delta^+\)-\gls{sph} and demonstrated this on an extended set of test problems.
Subsequently, \citet{chironFastAccurateSPH2019} proposed further improvements to the method of \citet{barcaroloAdaptiveParticleRefinement2014}.
Most importantly, \citet{chironFastAccurateSPH2019} introduced a buffer region of guard particles at the refinement interface to avoid the interaction of differently sized particles.
Again, this was implemented atop \(\delta^+\)-\gls{sph} to demonstrate application on an extended set of test problems by \citet{sunMultiresolutionDeltaSPHTensile2018}.

\citet{mutaEfficientAccurateAdaptive2022} proposed a new efficient and accurate adaptivity procedure.
Along with their own improvements, they integrated concepts from \citet{feldmanDynamicRefinementBoundary2007,vacondioSmoothedParticleHydrodynamics2012,vacondioVariableResolutionSPH2013,yangAdaptiveResolutionMultiphase2019} to build a comprehensive adaptivity procedure.
They made use of non-interacting background particles in their procedure to automatically define the refinement regions.
Subsequently, this procedure was updated to eliminate the requirement of background particles, thereby enhancing its efficiency~\citep{haftuParallelAdaptiveWeaklycompressible2022}. Recently, \citet{mutaSecondOrderVariableResolutionWeaklyCompressible2024} have proposed the initialization the offspring particles with properties linearly extrapolated from the parent to improve this procedure further.
By implementing the adaptivity procedures atop established second order convergent building blocks~\citep{ramachandranEntropicallyDampedArtificial2019, negiHowTrainYour2022a,negiTechniquesSecondorderConvergent2022}, \citet{mutaSecondOrderVariableResolutionWeaklyCompressible2024} were able to demonstrate second order convergence for the adaptivity procedure using \gls{mms}~\citep{negiHowTrainYour2021}.

Deviating from the conventional practice of considering mass as the basis for refinement and derefinement, \citet{sunAccurateSPHVolume2021} proposed an adaptivity procedure based on the particles' associated volumes.
Since density can vary in compressible flows, mass-based criteria cannot be used to define resolutions, and the usage of associated particle volumes as the basis for refinement and refinement becomes indispensable.
\citet{sunAccurateSPHVolume2021} have demonstrated the application of their adaptivity procedure on compressible multiphase problems.
However, in their work, adaptivity is employed merely to target homogenous and isotropic particle distributions; they do not consider refining particles near shocks or other regions of interest.

Most works that have been described thus far consider pairwise merging of particles during derefinement.
Some exceptions to these are \citet{xiongGPUacceleratedAdaptiveParticle2013, sprengLocalAdaptiveDiscretization2014,sprengAdvancedStudyDiscretizationerrorbased2020, havasi-tothParticleCoalescingAngular2020, sunConservativeParticleSplitting2023}.
Of these, the recent works of \citet{havasi-tothParticleCoalescingAngular2020} and \citet{sunConservativeParticleSplitting2023} offer novel approaches that maximize conservation.
Most of the works mentioned above also discuss and demonstrate adaptivity in only 2D.
A few exceptions include the works of \citet{vacondioVariableResolutionSPH2016,sunMultiresolutionDeltaSPHTensile2018,sprengAdvancedStudyDiscretizationerrorbased2020}. 

The adaptivity procedure that we propose in this work is primarily derived from the works of \citet{sunAccurateSPHVolume2021} and \citet{haftuParallelAdaptiveWeaklycompressible2022}.
Before we delve into the details of the adaptivity procedure, we provide a brief overview of the underlying scheme on which the adaptivity procedure is demonstrated.
 
\section{Methodology} \label{sec:formulation}

\subsection{Governing equations} \label{sec:governing-eq}

We consider an inviscid, compressible fluid in this work.
This is governed by the Euler equations, written as
\begin{equation}
	\label{eq:continuity} \odv{\rho}{t} =-\rho \ \nabla \cdot \boldsymbol{u},
\end{equation}
\begin{equation}
	\label{eq:momentum} \odv{\boldsymbol{u}}{t} =-\frac{1}{\rho}\ \nabla p,
\end{equation}
\begin{equation}
	\label{eq:energy} \odv{e}{t} =-\frac{p}{\rho}\ \nabla \cdot \boldsymbol{u}.
\end{equation}
Here, \(\rho\) is the density, \(\boldsymbol{u}\) is the velocity, \(p\) is the pressure, and \(e\) is the thermal energy per unit mass.
\(\mathrm{d}( \cdot ) / \mathrm{d} t\) represents the material derivative. This system is closed with an equation of state
\begin{equation}
	\label{eq:igeos} p=(\gamma-1) \rho e ,
\end{equation}
with the ideal gas assumption.
Here, \(\gamma\) is the ratio of the specific heat at constant pressure to constant volume.
\(\gamma\) is constant for a calorifically perfect gas.

\Cref{eq:continuity,eq:momentum,eq:energy} form a Lagrangian specification.
In a discrete sense, this would imply that each point follows a fluid element as it moves. 
If we were to consider points moving with a transport velocity \(\tilde{\boldsymbol{u}}\) instead of \(\boldsymbol{u}\), the equations would read
\begin{equation}
	\label{eq:continuity-tvf} \odv[style-inf-num=\mathrm{\tilde{d}}]{\rho}{t} =-\rho\nabla\cdot \boldsymbol{u}-\rho\nabla\cdot \delta\boldsymbol{u} +\nabla\cdot \left(\rho \delta\boldsymbol{u}\right),
\end{equation}
\begin{equation}
	\label{eq:momentum-tvf} \odv[style-inf-num=\mathrm{\tilde{d}}]{\boldsymbol{u}}{t}= -\frac{1}{\rho}\nabla p+\rho \nabla \cdot \left(\boldsymbol{u}\otimes \delta \boldsymbol{u} \right) - \rho_i \boldsymbol{u}_i \nabla \cdot \delta \boldsymbol{u},
\end{equation}
and
\begin{equation}
	\label{eq:energy-tvf} \odv[style-inf-num=\mathrm{\tilde{d}}]{e}{t}=-\frac{p}{\rho}\nabla \cdot \boldsymbol{u} - e \nabla\cdot \delta\boldsymbol{u} +\nabla\cdot \left(e \delta\boldsymbol{u}\right),
\end{equation}
where \(\delta \boldsymbol{u} = \boldsymbol{u} - \tilde{\boldsymbol{u}}\).
Here, \(\mathrm{\tilde{d}} (\cdot)/\mathrm{d} t\) represents the material derivative with respect to the transport velocity, \(\tilde{\boldsymbol{u}}\).
This is referred to as \gls{tvf} in \gls{sph} literature~\citep{TVFAdami2013}.
We refer the readers to \citet{adepuCorrectedTransportvelocityFormulation2023} for a detailed derivation of these equations.

\subsection{Discretized governing equations} \label{sec:discretized-eq}
We refer the readers to \citet{monaghanSmoothedParticleHydrodynamics2005,priceSmoothedParticleHydrodynamics2012,violeauFluidMechanicsSPH2015} for a detailed introduction to \gls{sph} discretization.
We omit the very basic details of \gls{sph} discretization and list the discretized forms of the governing equations here for notational clarity. 

The discretized version of the equation of state is not very different from the continuous form given by \cref{eq:igeos}.
The pressure at particle \(i\) is computed as
\begin{equation}
	p_i=(\gamma-1) \rho_i e_i .
\end{equation}

For the rest of the equations, we follow \citet{rosswogAgrangianHydrodynamicsCode2020,rosswogSimpleEntropybasedDissipation2020}. Several schemes are listed in the work of \citet{rosswogBoostingAccuracySPH2015,rosswogAgrangianHydrodynamicsCode2020}.
We consider the scheme named \gls{mi1} with a few modifications, as explained further.

The technique used by \citet{rosswogAgrangianHydrodynamicsCode2020} for updating the smoothing length, $h$, is not adopted in this study. 
Instead, equations
\begin{equation}
	h_i = h_{\text{fact}} \left(\frac{1}{n_i}\right)^{1/d}
\end{equation}
and
\begin{equation}
	n_i = \sum_j W_{i}
\end{equation}
are solved iteratively using a Newton-Raphson method to update the smoothing length~\citep{priceSmoothedParticleHydrodynamics2012,hopkinsGeneralClassLagrangian2013,puriComparisonSPHSchemes2014}.
Here, \(j \in \mathcal{N}\) where \(\mathcal{N}\) is the set of particles in the neighbourhood of particle \(i\).
\(W\) is the \gls{sph} kernel; where \(W_{i}\) stands for \(W(\left|\boldsymbol{r}_{i}-\boldsymbol{r}_{j}\right|, h_{i})\). \(\boldsymbol{r}\) represents the position vector. \(d\) is the dimensionality of the problem. \(h_{\text{fact}}\) is a constant set as 1.5. \(n\) is referred to as the number density.

Instead of employing a discretized continuity equation to update density, the \gls{mi1} scheme relies on computing density using the gather form~\citep{hernquistTREESPHUnificationSPH1989} of summation density,
\begin{equation}
	\label{eq:mi1-density} \rho_i = \sum_j m_j W_{i}.
\end{equation}

For discretization of \cref{eq:momentum} and \cref{eq:energy}, \gls{mi1} scheme avoids the use of analytical kernel gradient, \(\nabla W\).
The discrete form of \cref{eq:momentum} and \cref{eq:energy} are given as
\begin{equation}
	\label{eq:mi1-momentum}
	\odv{\boldsymbol{u}_i}{t} = -\sum_j m_j \left(\frac{p_i}{\rho_i^2} \boldsymbol{G}_i + \frac{p_j}{\rho_j^2} \boldsymbol{G}_j\right),
\end{equation}
and
\begin{equation}
	\label{eq:mi1-energy}
	\odv{\boldsymbol{e}_i}{t} = \frac{p_i}{\rho_i^2} \sum_j m_j \boldsymbol{u}_{ij} \cdot \boldsymbol{G}_i,
\end{equation}
respectively, where $\boldsymbol{u}_{ij} = \boldsymbol{u}_i - \boldsymbol{u}_j$,
\begin{equation}
	\label{eq:mi1-Gi}
	\boldsymbol{G}_i = \boldsymbol{C}_i \cdot \boldsymbol{r}_{ji} W_{i},
\end{equation}
and
\begin{equation}
	\label{eq:mi1-Gj}
	\boldsymbol{G}_j = \boldsymbol{C}_j \cdot \boldsymbol{r}_{ji} W_{j},
\end{equation}
where in turn
\begin{equation}
	\label{eq:mi1-C}
	\boldsymbol{C}_i = \left(\sum_j \boldsymbol{r}_{ji} \otimes \boldsymbol{r}_{ji}
	W_i \frac{m_j}{\rho_j}\right)^{-1}.
\end{equation}

With the addition of artificial viscosity and artificial conduction terms, equations \cref{eq:mi1-momentum} and \cref{eq:mi1-energy} become
\begin{equation}
	\label{eq:mi1-momentum-av}
	\odv{\boldsymbol{u}_i}{t} = -\sum_j m_j \left(\frac{p_i + q_i}{\rho_i^2} \boldsymbol{G}_i + \frac{p_j + q_j}{\rho_j^2} \boldsymbol{G}_j\right),
\end{equation}
and
\begin{equation}
	\label{eq:mi1-energy-ac}
	\odv{\boldsymbol{e}_i}{t} = \frac{p_i + q_i}{\rho_i^2} \sum_j m_j \boldsymbol{u}_{ij} \cdot \boldsymbol{G}_i - \alpha_e \sum_j v_{\text{sig}, e, ij} \left(\tilde{e}_i - \tilde{e}_j\right) \frac{\left|\boldsymbol{G}_i + \boldsymbol{G}_j \right|_2}{2} \frac{m_j}{\rho_{ij}},
\end{equation}
where
\begin{equation}
	q_i = \rho_i \left(-\alpha c_{i} \mu_i + \beta \mu^2 \right),
\end{equation}
\begin{equation}
	\mu_i = \min{\left(0, \frac{\check{\boldsymbol{u}}_{ij} \cdot \boldsymbol{\eta}_{ij}}{\left|\boldsymbol{\eta}_i\right|^2 + \epsilon^2}\right)}.
\end{equation}
and
\begin{equation}
	v_{\text{sig}, e, ij} = \sqrt{\frac{\left|p_i-p_j\right|}{\rho_{ij} }}.
\end{equation}
Here, \(\boldsymbol{\eta}_{ij} = \boldsymbol{r}_{i} / h_i\).
$c$ is the speed of sound, symmetrized as \(c_{ij} = (c_i + c_j)/2\) where \(c_i=\sqrt{\gamma p_i / \rho_i}\). \(\beta\), \(\epsilon\) and \(\alpha_e\) are constants set as 2, 0.1 and 0.05, respectively. \(\alpha\) allowed to vary with the use of a dissipation limiter. \(\check{\boldsymbol{u}}_{ij} = \check{\boldsymbol{u}}_i - \check{\boldsymbol{u}}_j\) where  \(\check{\boldsymbol{u}}_i\) and \( \check{\boldsymbol{u}}_j\) represent the velocity of particles \(i\) and \(j\) quadratically reconstructed to their mid-point, \(\left(\boldsymbol{r}_i + \boldsymbol{r}_j\right) / 2\).
This reconstruction can be expressed as
\begin{equation}
	\check{\boldsymbol{u}}_i = \boldsymbol{u}_i + \Phi_{ij} \left( \frac{1}{2}\langle\nabla \boldsymbol{u}\rangle_i \cdot \boldsymbol{r}_{ij} + \frac{1}{8} \langle\nabla^2 \boldsymbol{u}\rangle_i : \boldsymbol{r}_{ij} \otimes \boldsymbol{r}_{ij}\right),
\end{equation}
where \(\Phi_{ij}\) is a slope limiter. The angle brackets represent the \gls{sph} approximation of the operation, specifically gradient and Laplacian operations in the above expression. The slope limiter used by \citet{rosswogAgrangianHydrodynamicsCode2020} was given by \citet{vanleerUltimateConservativeDifference1974a} and modified for \gls{sph} by \citet{frontiereCRKSPHConservativeReproducing2017}. It reads
\begin{equation}
	\Phi_{i j}=
	\begin{cases}
		\max \left[0, \min \left[1, \frac{4 A_{i j}}{\left(1+A_{i j}\right)^2}\right]\right] & \text { if } \eta_{a b}>\eta_{\text {crit}} \\ \max \left[0, \min \left[1, \frac{4 A_{i j}}{\left(1+A_{i j}\right)^2}\right]\right] e^{-\left(\frac{\eta_{i j}-\eta_{\text {crit}}}{0.2}\right)^2} & \text {otherwise},
	\end{cases}
\end{equation}
where
\begin{equation}
	A_{i j}=\frac{\langle \nabla \boldsymbol{u}\rangle_i : \boldsymbol{r}_{i j} \otimes \boldsymbol{r}_{i j}}{\langle \nabla \boldsymbol{u}\rangle_j : \boldsymbol{r}_{i j} \otimes \boldsymbol{r}_{i j}},
\end{equation}
\begin{equation}
	\boldsymbol{\eta}_{i j}=\min{\left(\boldsymbol{\eta}_{i}, \boldsymbol{\eta}_{j}\right)},
\end{equation}
and \(\eta_{\text{crit}}\) = 0.3.

To facilitate penetration shield explained in \cref{sec:boundary-treatment}, \cref{eq:mi1-momentum-av} and \cref{eq:mi1-energy-ac} are augmented with \gls{tvf} terms as
\begin{equation}
	\label{eq:mi1-momentum-tvf} \odv[style-inf-num=\mathrm{\tilde{d}}]{\boldsymbol{u}_i}{t} = \odv{\boldsymbol{u}_i}{t} + \rho_i\langle \nabla \cdot \left(\boldsymbol{u}\otimes \delta \boldsymbol{u} \right) \rangle_{i} - \rho_i \boldsymbol{u}_i \langle \nabla \cdot \delta \boldsymbol{u} \rangle_i,
\end{equation}
and
\begin{equation}
	\label{eq:mi1-energy-tvf} \odv[style-inf-num=\mathrm{\tilde{d}}]{\boldsymbol{e}_i}{t} = \odv{\boldsymbol{e}_i}{t} - e_i \langle\nabla\cdot \delta\boldsymbol{u}\rangle_{i} +\langle\nabla\cdot \left(e \delta\boldsymbol{u}\right)\rangle_{i},
\end{equation}
respectively. Here,
\begin{equation}
	\langle\nabla \cdot \left(\boldsymbol{u}\otimes\delta\boldsymbol{u}\right)\rangle_{i}=\sum_j\left(\boldsymbol{u}_j\otimes\delta\boldsymbol{u}_j+\boldsymbol{u}_i\otimes\delta\boldsymbol{u}_i\right) \cdot \nabla_i W_{i j} \frac{m_j}{\rho_j},
\end{equation}
\begin{equation}
	\langle\nabla \cdot \left(e\delta\boldsymbol{u}\right)\rangle_{i}=\sum_j\left(e_j\delta\boldsymbol{u}_j+e_i\delta\boldsymbol{u}_i\right) \cdot \nabla_i W_{i j} \frac{m_j}{\rho_j},
\end{equation}
and
\begin{equation}
	\langle\nabla \cdot \delta \boldsymbol{u}\rangle_i = \sum_j \left(\boldsymbol{u}_j - \boldsymbol{u}_i\right) \cdot \nabla_i W_{i j} \frac{m_j}{\rho_j}.
\end{equation}

\subsection{Boundary treatment} \label{sec:boundary-treatment}
We use the boundary treatment method of \citet{adamiGeneralizedWallBoundary2012} with additional considerations for compressible flows prescribed by \citet{villodiRobustSolidBoundary2024}.
This method involves the use of ghost particles.
Properties like \(e, p\) and \(h\) are extrapolated from the fluid particles to the ghost particles as
\begin{equation}
	f_i=\frac{\sum_j f_j W_{i j}}{\sum_j W_{i j}}.
	\label{eq:adami-shepard}
\end{equation}
Note that the summation is over the neighbouring fluid particles instead of all the neighbouring particles.
Therefore, \(j\) excludes other ghost particles and represents only the fluid particles in the neighbourhood of the corresponding ghost particle \(i\).
Each component of the velocity is extrapolated using \cref{eq:adami-shepard} to obtain \(\boldsymbol{u}_{\text{extrapolated}}\).
The ghost particle is then assigned the velocity
\begin{equation}
	\label{eq:ghost-vel}
	\boldsymbol{u}_i = 2 \boldsymbol{u}_{i,\text{prescribed}} - \boldsymbol{{u}}_{i,\text{extrapolated}},
\end{equation}
where
\begin{equation}
	\label{eq:flip velocity normal interface} \boldsymbol{u}_{i,\text{prescribed}} = \boldsymbol{u}_{i,\text{interface}} + \boldsymbol{u}_{i,\text{extrapolated}} - \boldsymbol{u}_{i,\text{extrapolated}} \cdot \hat{\boldsymbol{n}}_i,
\end{equation}
Here, \(\hat{\boldsymbol{n}}\) is the normal to the interface.
\(\boldsymbol{u}\) is a notional velocity used just for computing the accelerations of the fluid particles. \(\boldsymbol{u}_{\text{interface}}\) is the velocity of the interface, i.e.,  \(\boldsymbol{u}_{i, \text{interface}}\) is the velocity with which the \(i^\text{th}\) ghost particle actually moves.
The pressure of the ghost particles is obtained using the extrapolated pressure, \(p_{\text{extrapolated}}\) as
\begin{equation}
	p_i = p_{i,\text{extrapolated}} + 2 \Delta s_{i,\text{g2i}} \left. \pdv{p}{n}\right|_i,
\end{equation}
where \(\Delta s_{i,\text{g2i}}\) is the shortest distance from the \(i^\text{th}\) ghost particle to the interface.
Exploiting \cref{eq:momentum}, \(\pdv{p}/{n}\) may be estimated as,
\begin{equation}
	\left.\pdv{p}{n}\right|_i = -\rho_i \boldsymbol{a}_i \cdot \hat{\boldsymbol{n}}_i,
\end{equation}
where \(\boldsymbol{a} = \odv{\boldsymbol{u}}/{t}\) is the acceleration of the ghost particle representing the accelerating interface.

The density of the ghost particles is obtained using the extrapolated \(e\) and computed \(p\) in the equation of state.
The mass \(m\) of the ghost particles is modified to ensure that the representative volume, \(m/\rho\), remains constant.

The shield proposed by \citet{villodiRobustSolidBoundary2024} is also applied.
If a fluid particle is on course to penetrate the interface, it is steered away using transport velocity as
\begin{equation}
	\label{eq:shield}
	\delta \boldsymbol{u}_i = \delta{u}_i \hat{\boldsymbol{n}}_{i,\text{in}}
\end{equation}
where
\begin{equation}
	\label{eq:shield-1}
	\delta u_i =
	\begin{cases}
		2.0 \frac{ \Delta s_{i} - \Delta s_{i,\text{f2g}} }{\Delta s_{i} } \hat{\boldsymbol{n}}_{i,\text{in}} \cdot \boldsymbol{u}_i & \text{if } \hat{\boldsymbol{n}}_{i,\text{in}} \cdot \boldsymbol{u}_i < 0 \text{ and } \Delta s_{i,\text{f2g}} < \Delta s_{i} \\ 0 & \text{otherwise}
	\end{cases}
\end{equation}
Here, \(\hat{\boldsymbol{n}}_\text{in}\) represents a unit vector that is normal to the interface, \(\Delta s_{i, \text{f2g}}\) is the distance from the \(i^\text{th}\) fluid particle to the nearest ghost particle, and \(\Delta s\) is the nominal spacing set as \((m/\rho)^{1/d}\).
\(\hat{\boldsymbol{n}}_\text{in}\) is updated before using it in \cref{eq:shield-1} by \gls{sph} interpolation of the normals, \(\hat{\boldsymbol{n}}\), carried by ghost particles.

\subsection{Time stepping and other parameters}

\gls{epec} integrator \citep{ramachandranEntropicallyDampedArtificial2019} is used for numerical integration. The time step is computed as
\begin{equation}
	\Delta t = C_{\text{CFL}}\min \left( \Delta t_{\text{vel}}, \Delta t_{\text{force}} \right),
\end{equation}
where
\begin{equation}
	\Delta t_{\text{vel}} = \frac{h_{\text{min}}}{\max \left( c \right)},
\end{equation}
\begin{equation}
	\Delta t_{\text{force}} = C_{\text{force}} \sqrt{\frac{h_{\text{min}}}{\max \left( \left| \odv{\boldsymbol{u}}{t} \right| \right)}}.
\end{equation}
$C_{\text{CFL}}$ and $C_{\text{force}}$ are constants.
Both are set as 0.5.

The Quintic Spline kernel~\citep{morrisModelingLowReynolds1997} is used for all the simulations in this study~\citep{negiTechniquesSecondorderConvergent2022}.
A fluid with \(\gamma = 1.4\) is used for all test problems, except for the oscillating piston chamber problem in \cref{sec:piston-chamber}.

\subsection{Adaptive Refinement} \label{sec:method}

The open-source, parallel implementation of \citet{haftuParallelAdaptiveWeaklycompressible2022} presents itself as an excellent starting point to build upon.
Meanwhile, volume-based adaptivity by \citet{sunAccurateSPHVolume2021} emerges closest to our objectives.
Therefore, we structure our implementation within the skeletal framework of the aforementioned open-source implementation.
We incorporate ideas from both \citet{haftuParallelAdaptiveWeaklycompressible2022} and \citet{sunAccurateSPHVolume2021} as summarized in \cref{tab:inspire}.
The overall procedure is shown in \cref{fig:adapt-flowchart}. 

\begin{table}
	\centering
	\begin{tabular}{m{.28\textwidth} c m{.40\textwidth}}
		\hline
		Feature                                         & Source                                              & Comments                                                         \\
		\hline \hline
		Splitting Pattern                               & \citet{sunAccurateSPHVolume2021}                    & Modified to ensure volume conservation
		\\ \hline
		Splitting Algorithm                             & \citet{haftuParallelAdaptiveWeaklycompressible2022} &
		Particle identification based on volume                                                                                                                                          \\ \hline
		Merging Algorithm                               & \citet{haftuParallelAdaptiveWeaklycompressible2022} & Identification based on volume \newline Support repeated merging \\ \hline
		Particle creation \& deletion                   & \citet{haftuParallelAdaptiveWeaklycompressible2022} & No changes                                                       \\ \hline
		Split \& Merge Threshold                        & \citet{sunAccurateSPHVolume2021}                    & No changes                                                       \\ \hline
		Regional Refinement \newline (Local Adaptivity) & \citet{haftuParallelAdaptiveWeaklycompressible2022} & Based on volume instead of mass                                 \\ \hline
		Automated Refinement Bands                      & \citet{haftuParallelAdaptiveWeaklycompressible2022} & Based on volume instead of mass                                  \\ \hline
		Solution Adaptivity                             & \citet{haftuParallelAdaptiveWeaklycompressible2022} & Resolve shocks instead of vortices                               \\
		\hline
	\end{tabular}
	\caption{Summary of features}
	\label{tab:inspire}
\end{table}

\begin{figure}
	\centering
	\scalebox{0.8}{
		\begin{tikzpicture}[node distance=1.5cm]
		\node (start) [startstop] {Start};
		\node (assignds) [process, below of=start] {Assign \(\Delta s\) for local and solution adaptivity};
		\node (updtvol) [process, below of=assignds] {Update volume bands};
		\node (marksplt) [process, below of=updtvol] {Mark split and merge worthy particles};
		\node (findmerge) [process, below left of=marksplt, yshift=-0.5cm, xshift=-2cm] {Find partners for merge worthy};
		\node (spltgrad) [process, below right of=marksplt, yshift=-0.5cm, xshift=2cm] {Compute \(\langle\nabla\rho\rangle, \langle\nabla e \rangle, \langle\nabla\boldsymbol{u}\rangle\) for split worthy};
		\node (procmerge) [process, below of=findmerge] {Merge particles};
		\node (addrem) [process, below=3.3cm of marksplt] {Add or delete particles};
		\node (initoffs) [process, below of=addrem, minimum width=7cm, text width=7cm] {Initialize offspring particles};
		\node (updtneigh) [process, below of=initoffs] {Update neighbour lists};
		\node (decision) [decision, below of=updtneigh, yshift=-0.6cm, aspect=1.8] {\(i_m < n_m\)?};
		\node (procmerge2) [process, right of=decision, xshift=5cm] {Merge particles};
		\node (findmerge2) [process, below of=procmerge2] {Mark merge worthy and find partners};
		\node (delete) [process, left of=decision, xshift=-4cm] {Delete particles};
		\node (updtneigh2) [process, below of=delete] {Update neighbour lists};
		\node (poshift) [process, below of=updtneigh2] {Position shift};
		\node (stop) [startstop, below of=decision, yshift=-1.5cm] {Stop};

		\draw[arrow] (start) -- (assignds);
		\draw[arrow] (assignds) -- (updtvol);
		\draw[arrow] (updtvol) -- (marksplt);
		\draw[arrow] (marksplt) |- (findmerge);
		\draw[arrow] (marksplt) |- (spltgrad);
		\draw[arrow] (findmerge) -- (procmerge);
		\draw[arrow] (procmerge) -| (addrem);
		\draw[arrow] (spltgrad.south) |- ([xshift=0.8cm]procmerge.east) -| (addrem.north);
		\draw[arrow] (addrem) -- (initoffs);
		\draw[arrow] (initoffs) -- (updtneigh);
		\draw[arrow] (updtneigh) -- node[anchor=west] {\(i_m = 0\)}(decision);
		\draw[arrow] (decision) |- node[anchor=south, xshift=1.5cm] {yes} (findmerge2);
		\draw[arrow] (findmerge2) -- (procmerge2);
		\draw[arrow] (procmerge2) --  node[anchor=south] {\(i_m+=1\)} (decision);
		\draw[arrow] (decision) -- node[anchor=south] {no}  (delete);
		\draw[arrow] (delete) -- (updtneigh2);
		\draw[arrow] (updtneigh2) -- (poshift);
		\draw[arrow] (poshift) -- (stop);
		\end{tikzpicture}
	}
	\caption{Flowchart of the adaptive refinement procedure}
	\label{fig:adapt-flowchart}
\end{figure}

\subsubsection{Splitting} \label{sec:splitting}

The splitting pattern is shown in \cref{fig:splitting-pattern}.
In practice, adaptive refinement is more useful in 2D and 3D.
Therefore, we do not pursue 1D adaptive refinement in this work.
The \(n_o\) offspring particles are placed at the vertices of a regular $d$-dimensional hypercube of size \(a\) centred at the parent particle.
\begin{equation}
	a = \frac{\sqrt[d]{V_p}}{n_o},
\end{equation}
where \(V = m / \rho\) and \(n_o = 2^d\). The volume of the parent particle is divided equally among the offspring particles
\begin{equation}
	V_o = \frac{V_p}{n_o}.
\end{equation}
\begin{figure}[H]
	\centering
	\begin{subfigure}[b]{0.3\textwidth}
		\centering
\begin{tikzpicture}[scale=2.0]
	\coordinate (O) at (0, 0);
	\coordinate (A) at (-0.5, 0.0);
	\coordinate (B) at (0.5, 0.0);

	\shade[ball color=gray] (O) circle (5pt);
	\draw[opacity=0.2] (A) -- (B);
	\foreach \point in {A, B}
		{
			\shade[ball color=cyan, opacity=0.8] (\point) circle (2.5pt);
		}
\end{tikzpicture}
		\caption{1D}
		\label{fig:splitting-pattern-1d}
	\end{subfigure}
	\begin{subfigure}[b]{0.3\textwidth}
		\centering
\begin{tikzpicture}[scale=2]
	\coordinate (O) at (0, 0);
	\coordinate (A) at (-0.5, -0.5);
	\coordinate (B) at (-0.5, 0.5);
	\coordinate (C) at (0.5, 0.5);
	\coordinate (D) at (0.5, -0.5);

	\shade[ball color=gray] (O) circle (5pt);
	\draw[opacity=0.2] (A) -- (B) -- (C) -- (D) -- cycle;
	\foreach \point in {A, B, C, D}
		{
			\shade[ball color=cyan, opacity=0.9] (\point) circle (2.5pt);
		}
\end{tikzpicture}
		\caption{2D}
		\label{fig:splitting-pattern-2d}
	\end{subfigure}
	\begin{subfigure}[b]{0.3\textwidth}
		\centering
\begin{tikzpicture}[scale=2]
	\coordinate (O) at (0, 0, 0);
	\coordinate (A) at (-0.5, 0.5, 0.5);
	\coordinate (B) at (-0.5, -0.5, 0.5);
	\coordinate (C) at (0.5, -0.5, 0.5);
	\coordinate (D) at (0.5, 0.5, 0.5);
	\coordinate (E) at (-0.5, 0.5, -0.5);
	\coordinate (F) at (-0.5, -0.5, -0.5);
	\coordinate (G) at (0.5, -0.5, -0.5);
	\coordinate (H) at (0.5, 0.5, -0.5);
	\draw[opacity=0.2] (A) -- (B) -- (C) -- (D) -- cycle;
	\draw[opacity=0.2] (A) -- (E);
	\draw[opacity=0.2] (B) -- (F);
	\draw[opacity=0.2] (C) -- (G);
	\draw[opacity=0.2] (D) -- (H);
	\draw[opacity=0.2] (E) -- (F) -- (G) -- (H) -- cycle;
	\foreach \point in {A, B, C, D}
		{
			\shade[ball color=cyan, opacity=0.9] (\point) circle (2.5pt);
		}
	\shade[ball color=gray] (O) circle (5pt);
	\foreach \point in {E, F, G, H}
		{
			\shade[ball color=cyan, opacity=0.9] (\point) circle (2.5pt);
		}
\end{tikzpicture}
		\caption{3D}
		\label{fig:splitting-pattern-3d}
	\end{subfigure}
	\caption{Splitting pattern in 1D, 2D, and 3D. The parent particle is shown in black, and the offspring particles are shown in cyan. The the parent particle is not retainted after the formation of offspring particles.}
	\label{fig:splitting-pattern}
\end{figure}

Regarding the properties of offspring particles, one obvious choice is to distribute the mass and volume of the parent particle equally among the offspring particles.
Other properties like density, velocity, and thermal energy can then be copied over from the parent particle.
This ensures the conservation of mass, volume, linear momentum, angular momentum, thermal energy, and kinetic energy.
In fact, \citet{feldmanDynamicRefinementBoundary2007} have shown that the only possible splitting strategy to ensure the conservation of momentum and kinetic energy is to let the offspring particles have the same velocity as the parent particle.
However, \citet{sunAccurateSPHVolume2021} have used linear extrapolation to determine the density, velocity, and thermal energy of the offspring particles.
Though this is not strictly conservative, with their oscillating piston chamber test case, they have shown that the violation of conservation is negligible in practice.
Moreover, according to \citet{mutaSecondOrderVariableResolutionWeaklyCompressible2024}, linear extrapolation warrants second-order convergence.

In the volume adaptive scheme, explicit conservation for momentum and thermal energy can be achieved by applying a correction after linear extrapolation.
Let \(\boldsymbol{r}_{o,k}\) be the position of the \(k\)\textsuperscript{th} offspring particle and \(\boldsymbol{r}_p\) be the position of the parent particle. 
Note that the \(p\) in the subscript denotes the parent particle, not to be confused with the pressure.
The density of the \(k\)\textsuperscript{th} offspring particle is obtained by extrapolating the properties of the parent particle as
\begin{equation}
	\rho_{o,k} = \rho_p + \langle \nabla \rho \rangle_p \cdot \left(\boldsymbol{r}_{o,k} - \boldsymbol{r}_p\right).
\end{equation}
Now, the mass of the offspring particle can be obtained as
\begin{equation} \label{eq:mass-offspring}
	m_{o,k} = \rho_{o,k}
	V_o.
\end{equation}

The velocity and specific thermal energy of the offspring particle are obtained as
\begin{equation}
	\label{eq:split-velocity}
	\boldsymbol{u}_{o,k} = \frac{m^*_o}{m_{o,k}}\left[\boldsymbol{u}_p + \langle \nabla \boldsymbol{u} \rangle_p \cdot \left(\boldsymbol{r}_{o,k} - \boldsymbol{r}_p\right)\right],
\end{equation}
and
\begin{equation}
	\label{eq:split-energy}
	e_{o,k} = \frac{m^*_o}{m_{o,k}}\left[e_p + \langle \nabla e \rangle_p \cdot \left(\boldsymbol{r}_{o,k} - \boldsymbol{r}_p\right)\right],
\end{equation}
where \(m^*_o\) be the mass of the offspring particle if the mass of the parent particle were distributed equally among the offspring particles, i.e,
\begin{equation}
	m^*_o = \frac{m_p}{n_o}.
\end{equation}
Finally, the pressure of the offspring particle is obtained using the equation of state.
As mass was not distributed equally among the offspring particles, factor \(m^*_o/m_o\) is crucial in \cref{eq:split-velocity,eq:split-energy} to ensure the conservation of momentum and energy.

It might appear that determining mass from density and volume as given by \cref{eq:mass-offspring} instead of distributing the mass of the parent particle equally among the offspring particles would violate the conservation of mass.
The placement of the offspring particles and the fact that volume is distributed equally along with the density being linearly extrapolated ensures that the mass of the offspring particles is actually conserved. This can be expressed as
\begin{equation}
	\sum_{k=1}^{n_o} m_{o,k} = \sum_{k=1}^{n_o} \rho_{o,k} V_o = \sum_{k=1}^{n_o} \left(\rho_p + \langle \nabla \rho \rangle_p \cdot \left(\boldsymbol{r}_{o,k} - \boldsymbol{r}_p\right)\right) V_o = m_p.
\end{equation}
Here, terms involving \(\langle \nabla \rho \rangle_p \) cancel out when summed over all offspring particles because of the symmetry in the placement of the offspring particles. Similarly, it is easy to verify that
\begin{equation}
	\sum_{k=1}^{n_o} m_{o,k} \boldsymbol{u}_{o,k} = m_p \boldsymbol{u}_p,
\end{equation}
and
\begin{equation}
	\sum_{k=1}^{n_o} m_{o,k} e_{o,k} = m_p e_p.
\end{equation}

Unlike the splitting scheme of \citet{sunAccurateSPHVolume2021} where volume is not conserved, equally distributing the volume of the parent particle among the offspring particles and incorporating this additional correction accords conservation of volume along with other properties.
It may be noted that kinetic energy and angular momentum are not conserved. Conservation of these quantities may be realized by simply turning off the extrapolation.
As we use this corrected extrapolation in the present work, a formal study to verify the convergence of the entire scheme is not beyond the broad scope of this study.
However, at this point, we mark this as a topic for future work and focus on the aspects mentioned in \cref{sec:intro}.

\subsubsection{Merging} \label{sec:merging} Merging is performed pairwise, i.e., two particles identified as each other's merge partners merge to form one.
This is simple and can be performed in parallel.
However, as a particle splits into 4 in 2D and 8 in 3D, the rate of refinement substantially exceeds the rate of derefinement.
Moreover, freshly formed offspring particles might immediately fulfil the merge criteria.
For this, merging is repeated \(n_m\) times as shown in \cref{fig:adapt-flowchart}.
The removal of particles marked for deletion and updation of neighbour lists are performed only after all the merge attempts are completed.

Let \(k\) be the particle formed by merging particles \(i\) and \(j\).
Particle \(k\) inherits the mass of particles \(i\) and \(j\),
\begin{equation}
	m_k = m_i + m_j.
\end{equation}
The density of particle \(k\) is computed as
\begin{equation}
	\rho_k = \frac{m_k \rho_i \rho_j}{m_i \rho_j + m_j \rho_i}.
\end{equation}
Smoothing length is computed as
\begin{equation}
	h_k = \sqrt[d]{h_i^d + h_j^d}.
\end{equation}
The position, velocity, and specific thermal energy of particle \(k\) are obtained by mass-weighted averaging of the properties of particles \(i\) and \(j\) as
\begin{equation}
	\boldsymbol{r}_k = \frac{m_i}{m_k} \boldsymbol{r}_i + \frac{m_j}{m_k} \boldsymbol{r}_j,
\end{equation}
\begin{equation}
	\boldsymbol{u}_k = \frac{m_i}{m_k} \boldsymbol{u}_i + \frac{m_j}{m_k} \boldsymbol{u}_j,
\end{equation}
\begin{equation}
	e_k = \frac{m_i}{m_k} e_i + \frac{m_j}{m_k} e_j.
\end{equation}
Finally, the pressure of particle \(k\) is obtained using the equation of state.
In practice, one of the particles is assigned the properties of the merged particle, and the other is marked for deletion. To make this deterministic, the particle with the lower index is chosen to be the merged particle and the other is marked for deletion.

\subsubsection{Local adaptivity and solution adaptivity} \label{sec:spatial-solution-adaptivity}

Particles are initialized with the maximum expected reference spacing, \(\Delta s_i = \Delta s_{\max}\).
We define another constant, \(\Delta s_{\min}\) as the minimum expected spacing in the refined regions.
For local adaptivity, \(\Delta s\) can be set to \(\Delta s_{\min}\) for the particles that lie in the vicinity of a body or within a certain predefined region.
In a similar vein, solution adaptivity can be achieved by assigning a low \(\Delta s\) value to particles that satisfy some predefined criteria based on the solution at the time.
Thus, \(\Delta s_{\max} / \Delta s_{\min} > 1\) is required for local and volume adaptivity.
\(\Delta s_{\max} / \Delta s_{\min} = 1\) implies volume adaptivity in pursuit of uniform resolution like \cite{sunAccurateSPHVolume2021}, without solution adaptivity or local adaptivity.

\begin{figure}
    \centering
    \includegraphics[width=0.95\textwidth]{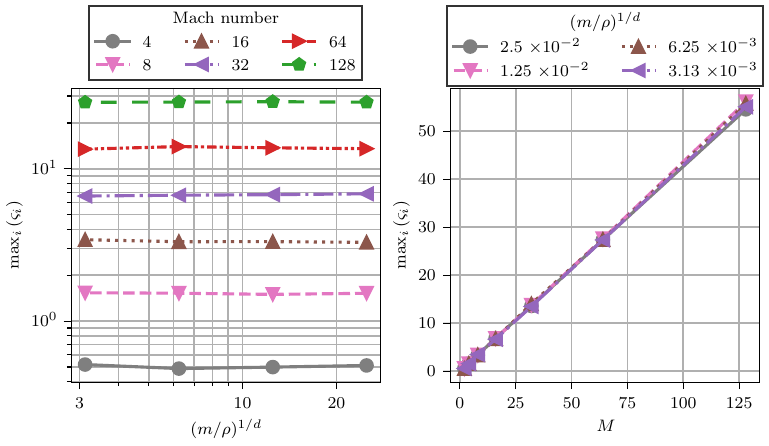}
    \caption{Variation of \(\max_i\left(\varsigma_i\right)\) with \((m/\rho)^{1/d}\) and $M$ for \(h = 1.5\).}
    \label{fig:indicat}
\end{figure}

\citet{mutaEfficientAccurateAdaptive2022,haftuParallelAdaptiveWeaklycompressible2022} demonstrated solution adaptivity using a vorticity-based criterion.
In this work, since we are considering compressible flows, we define a shock-based criterion for sharper shock resolution.
The presence of shock in a region is associated with a negative divergence of velocity.
This is exploited to identify particles near shocked regions.
Firstly, we define a quantity \(\varsigma\) for each particle as
\begin{equation}
	\label{eq:varsigma}
	\varsigma_i = -h_i  \langle\nabla \cdot \boldsymbol{u}\rangle_i.
\end{equation}
To justify this quantity, a parametric study was conducted by varying the Mach number, smoothing length and resolution using a 2D shock tube setup. The left and right states were assigned to set up shocks of different Mach numbers. \(\varsigma_i\) was noted after evolving the shock for a fixed distance. The variation of \(\max_i\left(\varsigma_i\right)\) with \((m/\rho)^{1/d}\) and Mach number for \(h_{\text{fact}} = 1.5\) is shown in \cref{fig:indicat}. This verifies that \(\varsigma_i\) is a reliable resolution-independent measure of the strength of the shock. Further, by plotting \(\varsigma\) as done in \cref{fig:ccor-explain}, one can observe that the band of particles with a high \(\varsigma\) is excessively narrow.
We aim to avoid the heavy churn due to refinement and derefinement very close to the shock. The same applies to particle shifting.
Therefore, to widen this band, we introduce \(\varsigma_s\) by inheriting the maximum \(\varsigma\) from among the neighbours
\begin{equation}
	\label{eq:varsigma-s}
	\varsigma_{s,i} = \max_j(\varsigma_{j}).
\end{equation}
With this, the reference spacing can be assigned as
\begin{equation}
	\label{eq:delta-s}
	\Delta s_i = 	\Delta s_{\min} \text{ if }  \varsigma_{s,i} > \varsigma_{rst},
\end{equation}
where \(\varsigma_{rst}\) is the threshold.
We find that \(\varsigma_{rst} = 0.4\) typically performs effectively but may also be adjusted according to the specific problem at hand.

If very fine refinements are required, following the above recipe for setting \(\Delta s\) may not be sufficient.
For such cases, one needs to create refinement bands that ensure a smoother transition between the fine and coarse regions.
A procedure to automate this is given by \citet{yangAdaptiveResolutionMultiphase2019,mutaEfficientAccurateAdaptive2022,haftuParallelAdaptiveWeaklycompressible2022}.
This procedure makes use of a parameter called refinement ratio, \(C_r\).
Refinement ratio can be defined as the ratio of the reference spacing between adjacent bands.
Let \(\Delta s_k\) be the reference spacing of the \(k\)\textsuperscript{th} band and \(\Delta s_{k+1}\) the reference spacing of the adjacent coarser \(k+1\)\textsuperscript{th} band.
Then, the refinement ratio would be
\begin{equation}
	C_r = \frac{\Delta s_{k+1}}{\Delta s_k}.
\end{equation}
We observe that a refinement ratio of 1.2 works well for most cases.
Once the refinement ratio is set, the algorithm from \citet{yangAdaptiveResolutionMultiphase2019, haftuParallelAdaptiveWeaklycompressible2022} can update the minimum and maximum expected volume, eventually ensuring a smoother transition between the refined and unrefined regions.

\subsubsection{Shock limited shifting} \label{sec:shock-limited-shifting}
The shifting magnitude is generally observed to be significant near shocks.
Even if we account for the shifting velocity through \gls{tvf} terms in the governing equations, we observe that the shocks are disturbed by shifting.
Moreover, discretization of the extra \gls{tvf} terms presented in \cref{sec:discretized-eq} and the shifting velocity is not antisymmetric, and thus, not conservative.
Therefore, shifting has to be restricted to smooth regions and be limited near shocks.

In the present work, we consider shifting proposed by \citet{michelParticleShiftingTechniques2022}.
We apply this not as a velocity but as a displacement.
This is applied within the adaptivity procedure as shown in \cref{fig:adapt-flowchart}.
The shifting displacement is expressed as
\begin{equation}
	\delta \boldsymbol{r}_i=0.5 \Delta t
	\begin{cases}
		0                                                                                                                           & \text{ if condition 1} ,      \\
		\mathcal{S}_{ij} \left(\frac{R}{W(\xi h_{ij}, h_{ij})}\right)^3 R \widehat{\nabla} C_i                                      & \text { else if condition 2}, \\
		\frac{\mathcal{S}_{ij}}{2} \frac{R}{W(\xi h_{ij}, h_{ij})} \frac{\widehat{\nabla} C_i}{\left\|\widehat{\nabla} C_i\right\|} & \text { otherwise, }
	\end{cases}
\end{equation}
where
\begin{equation}
	\mathcal{S}_{ij} = -\max_{j}\left(\left|\boldsymbol{u}_{ij} \cdot \frac{\boldsymbol{r}_{ij}}{\left\|\boldsymbol{r}_{ij}\right\|}\right|\right)
\end{equation}
and
\begin{equation}
	\widehat{\nabla}
	C_i = \sum_j \left[1 + R_m \left(\frac{W(\left|\boldsymbol{r}_{ij}\right|, h_{ij})}{W(\xi h_{ij}, h_{ij})}\right)^{R_n}\right] \nabla_i W_{ij} \frac{m_j}{\rho_j}.
\end{equation}
Conditions 1 and 2 are
\begin{equation}
	\varsigma_s < \varsigma_{st}
\end{equation}
and
\begin{equation}
	\left\|\left(\frac{R}{W(\xi h_{ij}, h_{ij})}\right)^3 R \widehat{\nabla}
	C_i\right\|<\frac{1}{2} \frac{R}{W(\xi h_{ij}, h_{ij})},
\end{equation}
respectively.
\(R\) is the kernel support radius.
For the quintic spline kernel, \(R = 3h\).
\(\varsigma_{st}\) is the threshold for \(\varsigma_s\) above which shifting is restricted.
We find that \(\varsigma_{st} = 0.4\) works well for most cases but may also be adjusted based on the problem at hand, just like \(\varsigma_{rst}\).
The constants \(R_m\) and \(R_n\) are set as 0.2 and 4, respectively~\citep{sunConsistentApproachParticle2019}.
\(\xi\) is the inflection point of the kernel; \(\xi =  0.759298480738450\) for the quintic spline~\citep{mutaEfficientAccurateAdaptive2022}.
Additionally, to prevent particles from moving too far, the shifting magnitude is capped to satisfy \(\left|\delta\boldsymbol{r}_i\right| \leq 0.25\left|\boldsymbol{u}_{i} \right|\Delta t\) and \(\left|\delta\boldsymbol{r}_i\right| \leq 0.25h_i\).

The shifting is more active in the regions of active refinement and derefinement.
Moreover, the shifting magnitude is much less than the displacement of the offspring from the parent particle after a split.
Thus, a Taylor correction is not applied after shifting, unlike the approach in  \cite{mutaEfficientAccurateAdaptive2022}.
This reduces the computational cost associated with estimating the gradients of the quantities required for the Taylor correction.
We find that this works well in practice; however, we do not discount the possible effects on convergence. Therefore, we mark this as a subject for future work regarding the formal convergence investigation.



\subsection{Other details}
The above scheme and the adaptivity procedure are implemented in \texttt{PySPH}~\citep{ramachandranPySPHPythonbasedFramework2021a}.
The particle packing algorithm of \citet{negiAlgorithmsUniformParticle2021} was used in creation of the ghost particles representing the bodies in the hypersonic cylinder~(\cref{sec:hcyl}), biconvex aerofoil~(\cref{sec:biconvex}), and Apollo reentry capsule~(\cref{sec:apollo}) problems.
The method of \citet{federicoSimulating2DOpenchannel2012b} implemented within \texttt{PySPH} by \citet{negiImprovedNonreflectingOutlet2020} is used for permeable boundaries.
The simulations were orchestrated using \texttt{automan}~\citep{ramachandranAutomanPythonBasedAutomation2018}.
\section{Results} \label{sec:results}

The proposed algorithm is applied to nine test problems, progressively uncovering various features of the proposed adaptivity procedure.
We demonstrate four modes of operation of the procedure,
\begin{enumerate}
	\item VA: Volume adaptivity,
	\item VA-SAS: Volume adaptivity with shock-aware shifting,
	\item VSA-SAS: VA-SAS with solution adaptivity, i.e., refinement near shocks,
	\item VSLA-SAS: VSA-SAS with local adaptivity, i.e., regional refinement around bodies.
\end{enumerate}
The first test problem in \cref{sec:dmr} is the double Mach reflection problem.
This problem demonstrates that incorporating simple adaptive resolution procedure results in the resolution of serious numerical instabilities that have been reported previously.
The next two problems demonstrate VA-SAS, targeting uniform resolution without solution adaptivity or local adaptivity.
The oscillating piston chamber test problem (\cref{sec:piston-chamber}) is a shockless two-dimensional problem used to validate the implementation.
Similar to the double Mach reflection problem in \cref{sec:dmr}, the shock diffraction problem in \cref{sec:diffraction} also demonstrates that adaptive refinement can remedy important issues commonly seen in compressible \gls{sph} simulations.
The compression corner problem in \cref{sec:compression-corner} is a simple problem that demonstrates VSA-SAS and serves to illustrate the working of the solution adaptivity and the shock-limited shifting.
The Noh's implosion problem in \cref{sec:noh} also demonstrates VSA-SAS and is used to illustrate the effectiveness of the proposed adaptivity procedure on moving shocks.
The hypersonic cylinder problem in \cref{sec:hcyl} and biconvex airfoil problem in \cref{sec:biconvex} demonstrate VSLA-SAS, i.e., solution adaptivity near shocks and local adaptivity around bodies in conjunction with the shock limited shifting.
The rotating square projectile problem in \cref{sec:rotsq} also demonstrates VSLA-SAS, but in a more complex scenario with a moving body.
Finally, the Apollo reentry capsule problem in \cref{sec:apollo} is used to demonstrate that the proposed refinement and derefinement procedures work well in 3D.

\subsection{Oscillating piston chamber} \label{sec:piston-chamber}
This is a two-dimensional compressible but shockless problem introduced by \citet{sunAccurateSPHVolume2021} to showcase the strengths of their volume adaptivity procedure.
The setup consists of a \(\SI{2}{\meter} \times \SI{1}{\meter}\) rectangular chamber with an oscillating piston on the left and fixed walls on the other three sides, as shown in \cref{fig:pco-action}.
The particles are initialized with a nominal spacing of \((m/\rho)^{1/d} = \qty{0.01}{\meter}\).
The chamber is filled with a compressible fluid with \(\gamma=1.25\).

The purpose of this problem is to demonstrate that no regressions have been introduced, the proposed adaptivity procedure works well, and that our implementation is comparable to, if not better than that of \citet{sunAccurateSPHVolume2021} with our changes.
Thus, we do not use the \gls{mi1} discretization scheme described in \cref{sec:formulation}.
We use our implementation of \citeauthor{sunAccurateSPHVolume2021}'s discretization scheme~\citep{villodiRobustSolidBoundary2024}.
Since we also use shifting velocity from \citet{sunAccurateSPHVolume2021} for this problem, the position shift (\cref{sec:shock-limited-shifting}) within the adaptivity procedure (\cref{fig:adapt-flowchart}) is turned off.
For this problem, the energy equation is discarded, and instead of \cref{eq:igeos}, the following equation of state is used to compute the pressure,
\begin{equation}
	p = p_0 \left(\frac{\rho}{\rho_0}\right)^\gamma.
\end{equation}
Here, we have used initial pressure, \(p_0=\)\SI{1.5e4}{\pascal} and an initial density, \(\rho_0=\)\SI{1}{\kg\per\meter\cubed}.

The position, \(x_p\) of the oscillating piston can be described as
\begin{equation}
	x_p = 0.75 L  \left(1 -  \cos{\left(4 \pi \frac{t}{t_p}\right)}\right),
\end{equation}
where \(L=\SI{2}{\meter}\) is the length of the chamber and \(t_p=\SI{1}{\second}\) is the time period of oscillation. As the piston oscillates, the analytical expression for pressure variation in the chamber is
\begin{equation}
	\label{eq:pco-analytical-pressure}
	p = p_0 \left(\frac{2}{2-x_p}\right)^{\gamma}.
\end{equation}

The pressure obtained from the simulation is compared with that computed using \cref{eq:pco-analytical-pressure} in \cref{fig:pco-line}~(above).
The number of particles in the chamber is also plotted in \cref{fig:pco-line}~(below).
It can be observed that the pressure obtained from the simulation closely matches the analytical solution.
The variation of the number of particles in the chamber also matches the results from the work of \citet{sunAccurateSPHVolume2021}.

\IfFileExists{figs_cache_PistonChamber_pco_action.pdf}%
{
	\begin{figure}[H]
		\centering
		\includegraphics[width=0.9\textwidth]{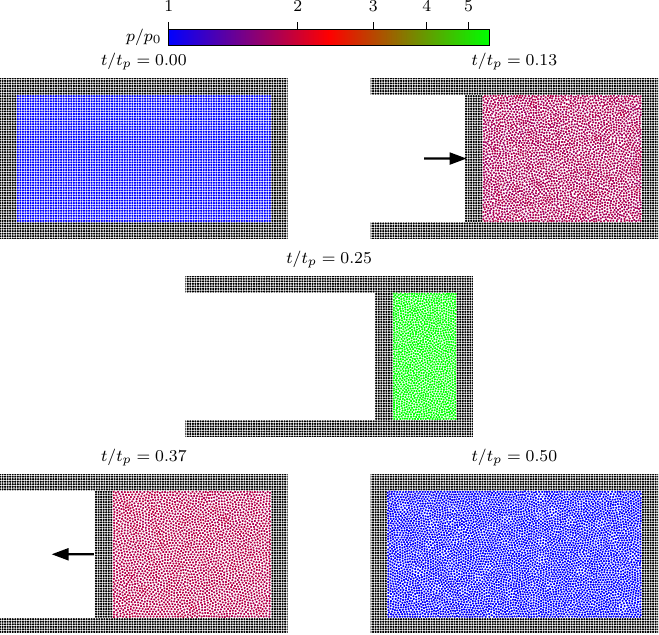}
		\caption{Particles in the chamber at different time instances coloured by pressure.}
		\label{fig:pco-action}
	\end{figure}
}{\textcolor{red}{Figure not found}}

\IfFileExists{figs_cache_PistonChamber_pco_line.pdf}%
{
	\begin{figure}[H]
		\centering
		\includegraphics[width=0.9\textwidth]{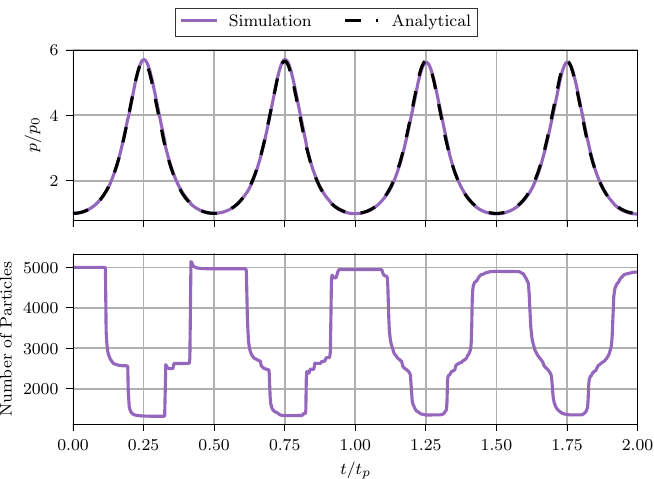}
		\caption{Variation of pressure and number of particles in the chamber with time}
		\label{fig:pco-line}
	\end{figure}
}%
{\textcolor{red}{Figure not found}}

\subsection{Double Mach reflection} \label{sec:dmr}
This problem involves simulation of the complex shock reflection structure that forms when a Mach 10 shock impinges a rigid wall at an angle of \(60^\circ\)~\citep{woodwardNumericalSimulationTwodimensional1984a}.
The setup described by \citet{tanInverseLaxWendroffProcedure2010} is used. Initially, a Mach 10 normal shock is positioned at the start of a \(\SI{60}{\degree}\) ramp. The inlet is $l_s = 1/6$ m away from the ramp. 
The undisturbed fluid upstream of the shock has a density of \(\rho_0=\)\qty{1.4}{\kilo\gram\per\cubic\meter} and a pressure of \(p_0=\)\qty{1}{\pascal}.
The simulation is run till $t_f=\qty{0.2}{\second}$.

This problem has only been attempted by \citet{gaoNewSmoothedParticle2023, wangEfficientTruncationScheme2024} and \citet{villodiRobustSolidBoundary2024} with \gls{sph}, to be best of our knowledge.
Of these, the work of \citet{gaoNewSmoothedParticle2023, wangEfficientTruncationScheme2024} deal with Eulerian-\gls{sph}, and therefore, does not encounter the challenges that we describe further.
At Mach 10, the change in density, and consequently the particle volume, across the shock is significant when simulated without adaptive resolution.
Moreover, fluid particles would need to interact with ghost particles that have a very different spacing.
These result in kinked Mach stem and particle clumping, leading to the formation of spot-like structures, as observed in \cref{fig:dmr-rho-noadapt}. The nominal particle spacing in this figure is \((m/\rho)^{1/d} = \qty{1.25e-2}{\meter}\).
Considerable noise around the primary slip line and some near-wall disturbance were also noted.
These artefacts would lead to blow-up if we increase the resolution, thus preventing us from using finer resolution for such problems.
\Cref{fig:dmr-rho-adapt} shows the final density distribution obtained with PyFR~\citep{witherdenPyFROpenSource2014} and with \gls{sph}. The resolution with PyFR is similar. The \gls{sph} result is obtained \((m/\rho)^{1/d} = \qty{1.5625e-3}{\meter}\) in VA mode, i.e., just with volume adaptivity, without additional local adaptivity or solution adaptivity. The result illustrates that with the application of the adaptivity procedure, major artefacts like clumping and kinked Mach stem are resolved. Other issues, such as the presence of noise, have been remedied to a large extent. The figure also demonstrates that we can go to finer resolution without any issues, allowing us to capture the Kelvin-Helmholtz formation on the primary slip line. 
While the adaptivity is effective in resolving the kinked Mach stem and clumping artefacts, the post-shock density overshoot is an artefact attributable to the discretization scheme used therein.
This is remedied by switching over to \gls{mi1} based discretization scheme, presented in \cref{sec:formulation}.
A major difference between the two discretization schemes is that the former uses a discrete approximation of the continuity equation while the latter uses summation density.
In a similar vein, \citet{pearlFSISPHSPHFormulation2022} switches to summation density for gases and discretized continuity for liquids and solids.
Alternatively, in-simulation periodic hybrid reinitialisation has also been reported in literature~\citep{colagrossiNumericalSimulationInterfacial2003, yangSimulationLiquidDrop2017,yangSmoothedParticleHydrodynamics2021}.

%


\IfFileExists{figs_cache_DoubleMachReflection_rho_noadapt.pdf}{
	\begin{figure}[H]
		\centering
		\begin{tikzpicture}
			\node[inner sep=0pt] {\includegraphics[width=0.95\textwidth]{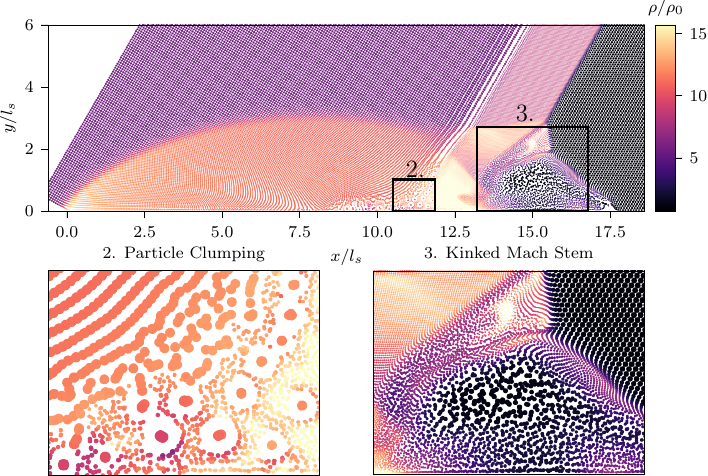}};
			\draw[thick,->] (3.45,3.2) to[out=45, in=0] (1.5,4.2) node[left] {\footnotesize 1. Post-shock density overshoot};
		\end{tikzpicture}
		\caption{Double Mach reflection result without adaptive resolution from the work of \citet{villodiRobustSolidBoundary2024}}
		\label{fig:dmr-rho-noadapt}
	\end{figure}
}{\textcolor{red}{Figure not found}}

\IfFileExists{figs_cache_DoubleMachReflection_rho_adapt.pdf}{
	\begin{figure}[H]
		\centering
		\includegraphics[width=0.95\textwidth]{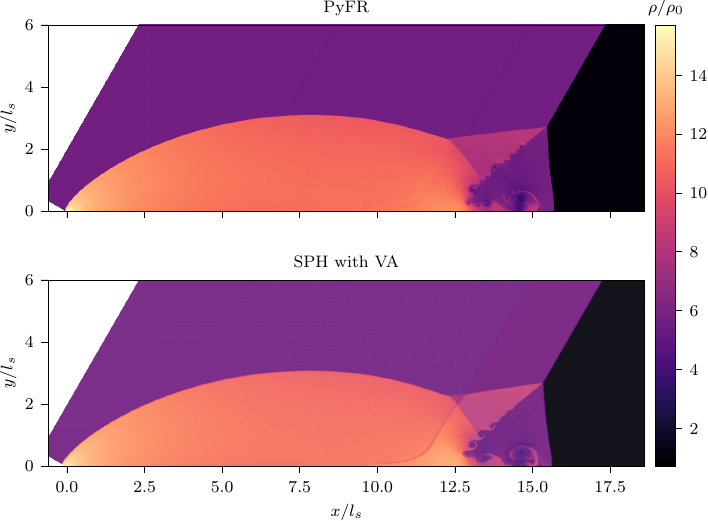};
		\caption{Double Mach reflection result using PyFR on top and \gls{sph} with VA on bottom.\
		}
		\label{fig:dmr-rho-adapt}
	\end{figure}
}{\textcolor{red}{Figure not found}}

%


\subsection{Shock Diffraction} \label{sec:diffraction}
Shock wave diffraction occurs when a travelling shock passes through an area of sudden expansion~\citep{gnaniExperimentalInvestigationShock2014}.
Initially, the normal shock at the tip of a backward-facing step as shown in \cref{fig:diffraction-mass}~(left).
The step height is \(l_s = \qty{0.625}{\meter}\).
The Mach 3 shock moves into stationary fluid with \(\rho_0=\qty{1.4}{\kilogram\per\metre\cubed}\) and \(p_0=\qty{1}{\pascal}\). Nominal particle spacing is \((m/\rho)^{1/d} = \qty{5e-3}{\meter}\).
The simulation is run till \(t_f=\SI{0.25}{\second}\).

Shock wave diffraction around corners is well studied experimentally~\citep{skewsShapeDiffractingShock1967,skewsPerturbedRegionDiffracting1967,bazhenovaUnsteadyInteractionsShock1984} and numerically~\citep{takayamaShockWaveDiffraction1991, sunVorticityProductionShock2003,fedorovNumericalStudyShockwave2008}.
Though this problem is considered as an established test case for validating compressible flow codes~\citep{ripleyNumericalSimulationShock2006,huangCuresNumericalShock2011,hennemannProvablyEntropyStable2021}, it has not been demonstrated to have been solved using \gls{sph} yet, to the best of our knowledge.
The rapid expansion ahead of the backward-facing step creates a low-density region.
This renders it a challenging problem for non-Eulerian \gls{sph}, where local particle density is representative of the local fluid density.
Therefore, adaptive resolution becomes absolutely essential.

The double Mach reflection in \cref{sec:dmr} was run in VA mode.
However, with the shock diffraction problem, redistribution the split particles in the low-density region becomes necessary. 
Therefore, shifting becomes unavoidable.
Hence, the shock diffraction problem is well-suited for demonstrating the VA-SAS mode.

The resolution deficiency, particle clumping and post-shock density overshoot were clearly observed when this problem was simulated using the discretization scheme from \citet{sunAccurateSPHVolume2021} without adaptive resolution.
The shock front is plagued by particle clumping, and the roll-up region is poorly resolved due to a deficiency in resolution.
These have been highlighted in \cref{fig:diffraction-rho-noadapt}.
We have also shown the final pressure field results using Eilmer~\citep{gibbonsEilmerOpensourceMultiphysics2023} and \gls{sph} with VA-SAS on the right in \cref{fig:diffraction-rho}. The \gls{sph} result clearly demonstrates that the proposed adaptive resolution procedure can resolve the highlighted issues effectively.
The initial and final mass distributions are shown in \cref{fig:diffraction-mass}.
The low-mass particles in the low-density region are a result of repeated splitting.
The comparison of resources required for this problem with \gls{mi1} based discretization scheme described in \cref{sec:formulation}, with and without adaptive resolution, is presented in \cref{tab:diffraction-comparison}.
This clearly highlights the proposed adaptive resolution procedure is effective in reducing computational resource usage and also expediting the simulation.

\begin{table}[H]
	\centering
	\begin{tabular}
		{|c|>{\centering\arraybackslash}p{0.15\textwidth}|>{\centering\arraybackslash}p{0.15\textwidth}|c|}
		\hline
		\multirow{2}{*}{Metric} & \multicolumn{2}{c|}{VA-SAS} & \multirow{2}{*}{Change}                     \\ \cline{2-3}
		                        & Off                              & On                     &                   \\ \hline \hline
		Final particles         & 91675                           & 68185                   & 25.7\% reduction  \\ \hline
		Time steps               & 458                             & 341                     & 25.6\% reduction  \\ \hline
		Simulation time (s)     & 170.11                          & 91.77                   & 1.87 times faster \\ \hline
	\end{tabular}
	\caption{Comparison of metrics with and without adaptive resolution for the shock diffraction problem.}
	\label{tab:diffraction-comparison}
\end{table}

\IfFileExists{figs_cache_Diffraction_rho_noadapt.pdf}{
	\begin{figure}[H]
		\centering
		\begin{tikzpicture}
			\node[](A) {\includegraphics[width=0.45\textwidth]{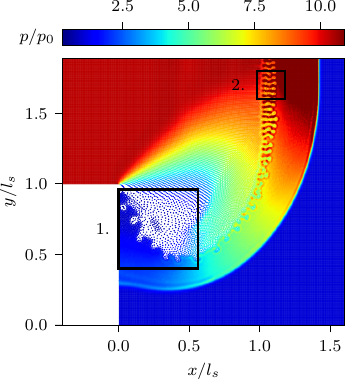}};
			\node[right=of A](B) {\includegraphics[width=0.35\textwidth]{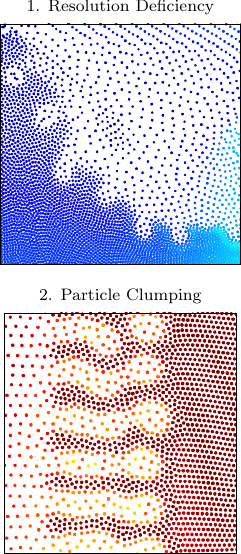}};
		\end{tikzpicture}
		\caption{Shock diffraction result without adaptive resolution using the discretization scheme from \citet{villodiRobustSolidBoundary2024}.}
		\label{fig:diffraction-rho-noadapt}
	\end{figure}
}{\textcolor{red}{Figure not found}}

\IfFileExists{figs_cache_Diffraction_rho.pdf}{
	\begin{figure}[H]
		\centering
		\includegraphics[width=0.95\textwidth]{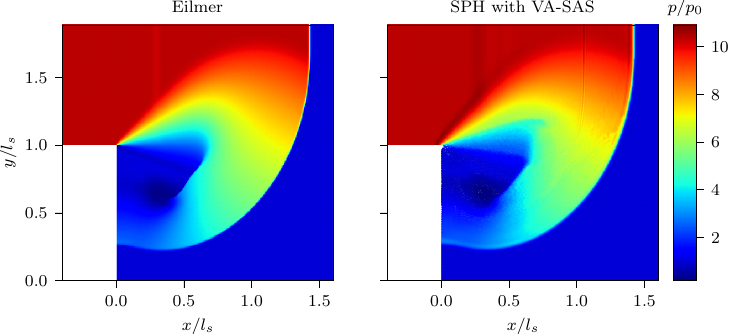}
		\caption{Shock diffraction using Eilmer on the left and SPH with VA-SAS on the right.}
		\label{fig:diffraction-rho}
	\end{figure}
}{\textcolor{red}{Figure not found}}

\IfFileExists{figs_cache_Diffraction_mass.pdf}{
	\begin{figure}[H]
		\centering
		\includegraphics[width=0.95\textwidth]{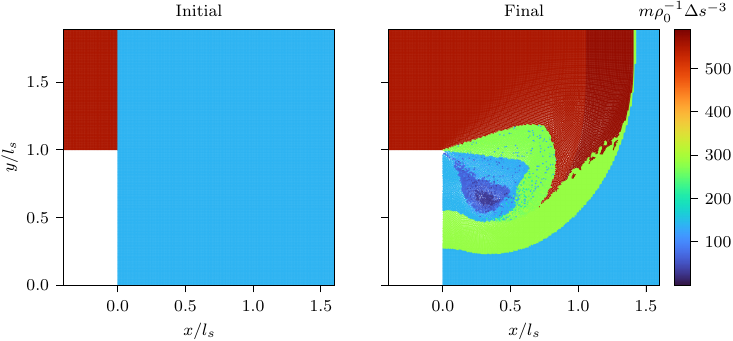}
		\caption{Shock diffraction result showing mass distribution.
			Initial mass distribution is shown on the left and final mass distribution is shown on the right.
		}
		\label{fig:diffraction-mass}
	\end{figure}
}{\textcolor{red}{Figure not found}}

Having demonstrated the VA and VA-SAS modes, we will now focus on the VSA-SAS mode in the next subsection.

\subsection{Compression Corner} \label{sec:compression-corner}
This test problem involves the computation of the supersonic flow field past a wedge.
Fluid with \(\gamma=1.4\) flowing over a wedge of half-angle $20^\circ$ at $M=3.0$ is considered.
The nominal spacing of particles at the inlet is \((m/\rho)^{1/d} =\qty{6.25e-2}{\meter}\). The wedge starts at origin and the inlet is \(l_s = \qty{1}{\metre}\) away from the wedge. The freestream pressure and density are \(p_\infty=\qty{1}{\pascal}\) and \(\rho_\infty=\qty{1.4}{\kilo\gram\per\cubic\meter}\), respectively. 
This problem is run till \(t_f=\qty{0.75}{\second}\).

With this simple case, we will try to illustrate the solution adaptivity mechanism in action.
The \(\varsigma\), \(\varsigma_s\), and \(\Delta s\) distributions can be seen in \cref{fig:ccor-explain}~(top left, top right and bottom left, respectively).
\(\varsigma\) is a shock-indicator. \(\varsigma_s\) is \(\varsigma\) spread using the equation \cref{eq:varsigma-s}.
When \(\varsigma_s\) is greater than the specified threshold, the particle is assigned \(\Delta s_i = \Delta s_{min}\).
As this propagates to the neighbouring particles by the algorithm mentioned in \cref{sec:spatial-solution-adaptivity}, bands of \(\Delta s\) are formed, as seen in \cref{fig:ccor-explain}~(bottom left).
Here, five \(\Delta s\) bands can be seen.
However, there are only two actual refinement levels since particles are not split or merged as long as they are within 2/3 to 8/5 of the reference volume, \(\left(\Delta s \right)^d\).
Particles are restricted from shifting near the shocks so that shocks are not disturbed.
The shifting restriction is also shown in the \cref{fig:ccor-explain}~(bottom right).
\Cref{fig:ccor-compare} illustrates the sharper change in the Mach number to the post-shock state in the adaptively resolved case.
The sharpness of the transition is also highlighted by the increased negative divergence of velocity in the adaptively resolved case.

\IfFileExists{figs_cache_Wedge_vol_adapt_explain.pdf}{
	\begin{figure}[H]
		\centering
		\includegraphics[width=0.9\textwidth]{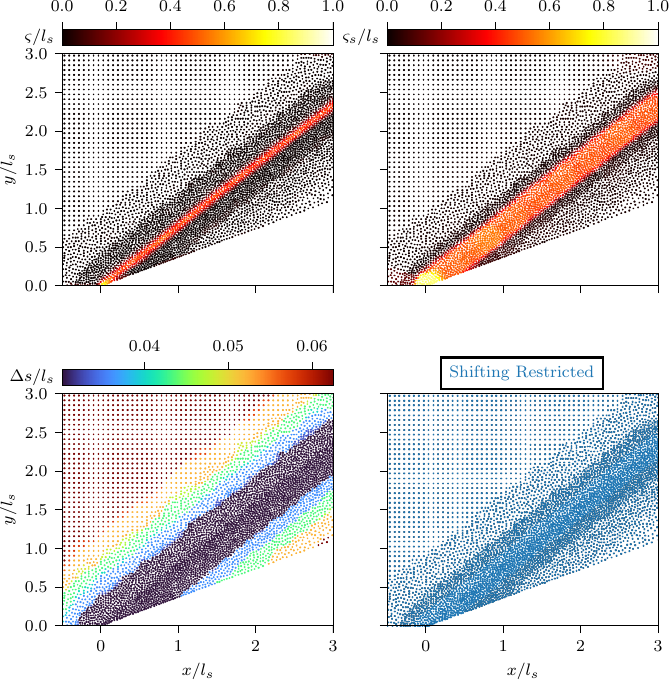}
		\caption{Result of Mach 3.0 flow over a compression corner showing the solution adaptivity mechanism.
			The plot on the top left shows particles coloured by \(\varsigma/l_s\), the plot on the top right shows particles coloured by \(\varsigma_s/l_s\), the plot on the bottom left shows particles coloured by \(\Delta s/l_s\), and the plot on the bottom right shows the particles with shifting restriction in blue.
		}
		\label{fig:ccor-explain}
	\end{figure}
}{\textcolor{red}{Figure not found}}

\IfFileExists{figs_cache_Wedge_vol_adapt_compare.pdf}{
	\begin{figure}[H]
		\centering
		\includegraphics[width=0.9\textwidth]{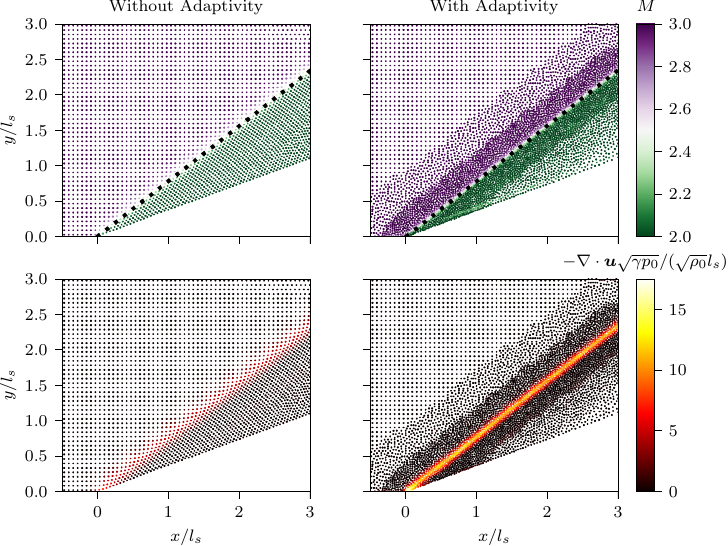}
		\caption{Results of Mach 3.0 flow over a compression corner.
			The left column shows the results without adaptivity, and the right column shows the results with adaptivity.
			The top row shows the Mach number, and the bottom row shows the negative divergence of velocity. In the plots on the top row, the theoretical location of the oblique shock is shown with a black dotted line.
		}
		\label{fig:ccor-compare}
	\end{figure}
}{\textcolor{red}{Figure not found}}

Having explained the solution adaptivity mechanism and demonstrated the VA-SAS mode, we will further demonstrate the VA-SAS mode on a more challenging problem with a moving shock in the next subsection.

\subsection{Noh's Implosion} \label{sec:noh}

We consider the cylindrical variant of Noh's Implosion~\citep{nohArtificialViscosityArtificial1983} problem.
This involves the implosion of fluid with \(\gamma=5/3\) and \(\rho_0=\qty{1.0}{\kilo\gram\per\meter\cubed}\) to a point at the centre of the domain.
Initially, the velocity is directed radially inward, towards the origin, \(u_{r,0} = \qty{1.0}{\meter\per\second}\). The nominal particle spacing at initialization is \((m/\rho)^{1/d} = \qty{2.10e-2}{\meter}\). A circular domain of radius \(l_s = \qty{1.25}{\meter}\) is used.
The simulation is run till \(t_f=\SI{0.50}{\second}\).

While the problem description is simple, this problem is recognized as challenging, not just for \gls{sph} but for other numerical methods as well~\citep{gehmeyrNohsConstantvelocityShock1997}. 
The analytical solution for this problem is well known and involves a strong shock wave that propagates outward from the centre.
The density behind the shock is 16 times the initial density, and the velocity behind the shock is zero.
This problem is used to assess the capability of the adaptive resolution procedure to handle strong shocks and large density variations.
We refer the readers to the works of \citet{hopkinsNewClassAccurate2015,rosswogSimpleEntropybasedDissipation2020,pearlFSISPHSPHFormulation2022} for observing the results of this problem with schemes incorporating recent advancements.

\Cref{fig:noh-spacing-ratio-6} demonstrates the working of the solution adaptivity procedure on moving shocks.
The results of various refinement settings are shown against the analytical solution in \cref{fig:noh-rho}.
Comparing \cref{fig:noh-rho}~(top-left vs top-right) and \cref{fig:noh-scatter}~(quadrant 1 vs quadrant 2), it is clear that implementing VA-SAS mode without careful consideration is suboptimal for this type of problem.
VA-SAS mode does reduce the number of particles required but also degrades the results.
This highlights the importance of oversight in applying the right adaptivity procedure.
One needs to carefully analyse the nature of the problem before applying the adaptivity procedure.

Comparing \cref{fig:noh-rho}~(top-left vs bottom-left) and \cref{fig:noh-scatter}~(quadrant 1 vs quadrant 3), we can see that VSA-SAS with \(\Delta s_{\max} / \Delta s_{\min}\) is able to improve the post-shock density and also make the shock sharper.
Comparing \cref{fig:noh-rho}(bottom-left vs bottom-right) and \cref{fig:noh-scatter}~(quadrant 3 vs quadrant 4), we can see that VSA-SAS with \(\Delta s_{\max} / \Delta s_{\min} = 6\) is able to improve the post-shock density further and also make the shock even sharper.
However, in \cref{fig:noh-rho}(bottom-right), we also observe some noise, especially right downstream of the shock.
This has been simulated at an increased resolution to further investigate the noise.
The noise does not grow uncontrollably and tends to remain bounded.
We conjecture that the source of noise lies upstream of the shock.
The noise is created as particles split and merge before the shock. 
This noise gets amplified as those particles pass through the shock. 
This is not an inherent issue with the adaptivity procedure itself.
Most results in the literature are obtained with particles arranged in a regular grid. With adaptivity, this clinical configuration of particles passing through the shock becomes impossible to achieve. 

With this test problem, we demonstrated VSA-SAS mode on a moving shock. In the next subsection, we will study the next test problem to demonstrate the VSLA-SAS mode.

\IfFileExists{figs_cache_Noh_spacing_ratio_6.0.pdf}{
	\begin{figure}[H]
		\centering
		\includegraphics[width=0.9\textwidth]{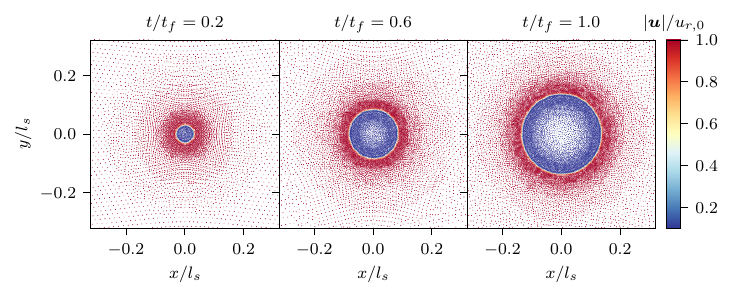}
		\caption{Evolution of particle distribution for Noh's implosion problem with solution adaptivity where \( \Delta s_{\max} / \Delta s_{\min} = 6 \).
			The left plot shows the particle distribution at \(t/t_f = 0.2\), the middle plot shows the particle distribution at \(t/t_f = 0.6\), and the right plot shows the particle distribution at \(t/t_f = 1.0\).
		}
		\label{fig:noh-spacing-ratio-6}
	\end{figure}
}{\textcolor{red}{Figure not found}}

\IfFileExists{figs_cache_Noh_rho.pdf}{
	\begin{figure}[H]
		\centering
		\includegraphics[width=0.9\textwidth]{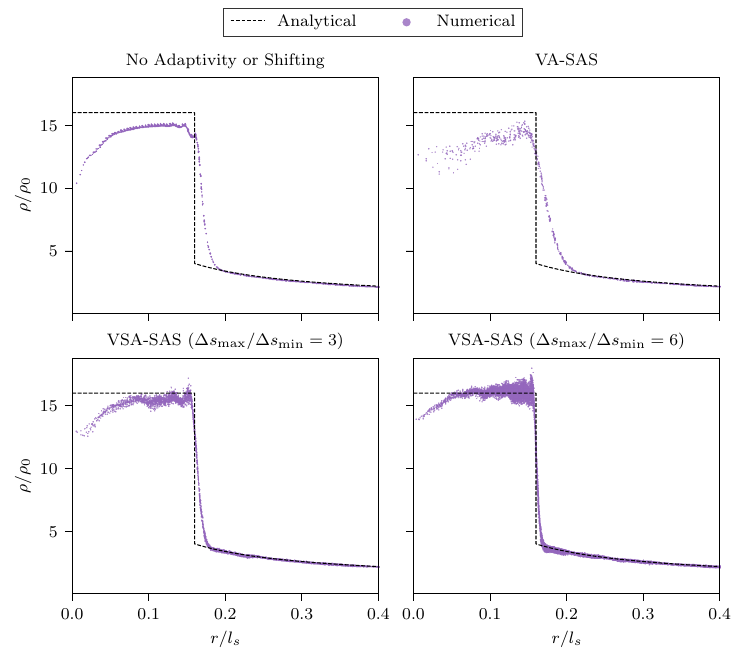}
		\caption{Resultant density distribution for Noh's implosion problem.
			The plot on the top left shows the results without adaptivity or shifting.
			The plot on the top right shows the results with basic volume adaptivity without solution adaptivity.
			The plot on the bottom left shows the results with solution adaptivity where \( \Delta s_{\max} / \Delta s_{\min} = 3 \).
			The plot on the bottom right also shows the results with solution adaptivity but with \( \Delta s_{\max} / \Delta s_{\min} = 6 \).
		}
		\label{fig:noh-rho}
	\end{figure}
}{\textcolor{red}{Figure not found}}

\IfFileExists{figs_cache_Noh_scatter.pdf}{
	\begin{figure}[H]
		\centering
		\includegraphics[width=0.45\textwidth]{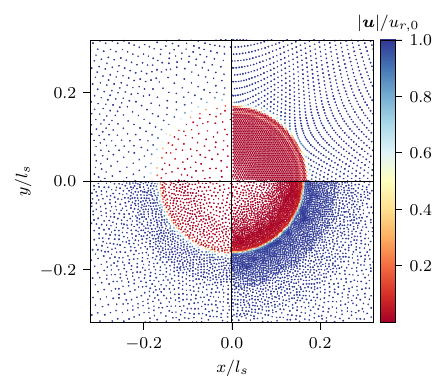}
		\caption{Comparison of particle distribution for Noh's implosion problem.
			The first quadrant displays the results without adaptivity or shifting.
			The second quadrant displays the results with basic volume adaptivity, without solution adaptivity.
			The third quadrant displays the results with solution adaptivity where \( \Delta s_{\max} / \Delta s_{\min} = 3 \).
			The fourth quadrant also displays the results with solution adaptivity, but with \( \Delta s_{\max} / \Delta s_{\min} = 6 \).
			The number of particles in quadrants one to four are 1963, 401, 1462 and 3479, respectively.
		}
		\label{fig:noh-scatter}
	\end{figure}
}{\textcolor{red}{Figure not found}}

\subsection{Hypersonic cylinder cone} \label{sec:hcyl}

This problem involves Mach 10 flow over the cylinder of radius \(l_s = \qty{1}{\meter}\) ahead of a $14.04^\circ$ cylinder cone.
The freestream pressure and density are \(p_\infty=\qty{1}{\pascal}\) and \(\rho_\infty=\qty{1.4}{\kilo\gram\per\cubic\meter}\), respectively. The nominal spacing of the incoming particles is \((m/\rho)^{1/d} = \qty{3.75e-2}{\meter}\). The simulation is run till \(t_f=\qty{3.5}{\second}\).

From the result presented in \cref{fig:hcyl-mach}, it can be observed that the particles around the cylinder are refined as part of the local adaptivity feature.
The solution adaptivity feature is demonstrated by the refinement of the particles in the bow shock region formed upstream of the cylinder cone.
This bow shock exhibits better agreement with the Billing's correlation~\citep{billigShockwaveShapesSphericaland1967} with VSLA-SAS than with no adaptivity and no shifting.

With this test problem on VSLA-SAS mode, we have demonstrated the local adaptivity and solution adaptivity features of the proposed adaptive resolution procedure. In the next subsection, we consider a test problem that highlights the potential of the VSLA-SAS mode in expediting the simulations and making the solution more accurate.

\IfFileExists{figs_cache_HypersonicCylinder_mach.pdf}{
	\begin{figure}[H]
		\centering
		\includegraphics[width=0.65\textwidth]{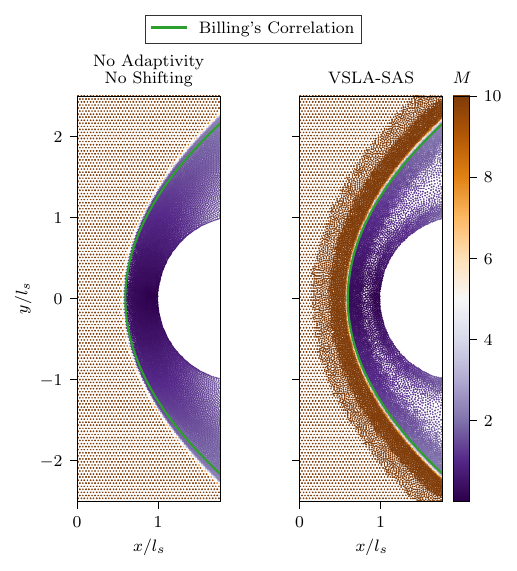}
		\caption{Hypersonic cylinder cone result with particles coloured by Mach number.}
		\label{fig:hcyl-mach}
	\end{figure}
}{\textcolor{red}{Figure not found}}

\subsection{Biconvex Airfoil} \label{sec:biconvex}
This problem involves Mach 4.04 flow over a biconvex airfoil at \(0^\circ\) angle of attack. The chord length of the airfoil is \(l_s = \qty{1}{\meter}\). 

Similar to the hypersonic cylinder cone problem in the previous subsection, this also serves as a demonstration of the local adaptivity and solution adaptivity capabilities of our proposed adaptive resolution method.
As depicted in \cref{fig:biconaer-mach}, the particles surrounding the airfoil undergo refinement as part of the local adaptivity feature.
The solution adaptivity feature is showcased by the refinement of particles around the shock.

In \cref{fig:biconpressure}, VA-SAS with \(\Delta s_{\min} = \Delta s_{\max} = \SI{1e-2}{\meter}\) and \(\Delta s_{\min} = \Delta s_{\max} = \SI{2.5e-3}{\meter}\) are compared against VSLA-SAS with \(\Delta s_{\max} = 4 \Delta s_{\min} = \SI{1e-2}{\meter}\).
The pressure distribution for VSLA-SAS with \(\Delta s_{\max} = 4 \Delta s_{\min} = \SI{1e-2}{\meter}\) is comparable to VA-SAS with \(\Delta s_{\min} = \Delta s_{\max} = \SI{2.5e-3}{\meter}\), and demonstrates good agreement with the inviscid theory. This clearly demonstrates the efficacy of local adaptivity and solution adaptivity. In \cref{tab:biconvex-comparison}, we have compared the resources required for VSLA-SAS with \(\Delta s_{\max} = 4 \Delta s_{\min} = \SI{1e-2}{\meter}\) against VA-SAS with \(\Delta s_{\min} = \Delta s_{\max} = \SI{2.5e-3}{\meter}\) and no adaptivity no shifting with particles of size $(m/\rho)^{1/2}=\SI{2.5e-3}{\meter}$ being introduced at inlet. 
It can be observed that VSLA-SAS with \(\Delta s_{\max} = 4 \Delta s_{\min} = \SI{1e-2}{\meter}\) requires much fewer particles and computational time. This clearly demonstrates the potential of the VSLA-SAS mode in making the solution more accurate and faster at the same time.

\IfFileExists{figs_cache_BiconvexAerofoil_mach.pdf}{
	\begin{figure}[H]
		\centering
		\includegraphics[width=0.95\textwidth]{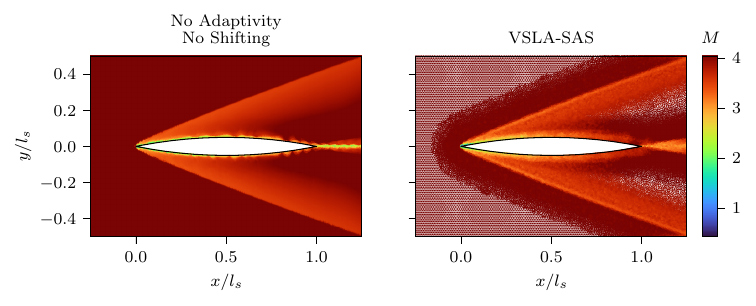}
		\caption{Biconvex airfoil results with particles coloured by Mach number.}
		\label{fig:biconaer-mach}
	\end{figure}
}{\textcolor{red}{Figure not found}}

\IfFileExists{figs_cache_BiconvexAerofoil_pressure.pdf}{
	\begin{figure}[H]
		\centering
		\includegraphics[width=0.65\textwidth]{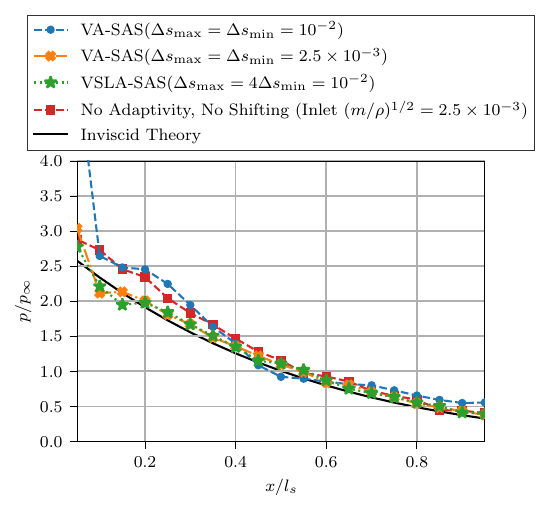}
		\caption{Pressure distribution over the biconvex airfoil.}
		\label{fig:biconpressure}
	\end{figure}
}{\textcolor{red}{Figure not found}}


\begin{table}[H]
	\centering
	\begin{tabular}		
		{|c|>{\centering\arraybackslash}p{0.20\textwidth}|c|c|}
		\hline
		Metric & No Adaptivity, No Shifting & VA-SAS & VSLA-SAS \\ \hline \hline
		Final particles    & 242,909 & 218,896 & 59,758 \\ \hline
		Timesteps          & 6,350   & 3,201   & 2,612  \\ \hline
		Simulation time (s)& 9,438   & 5,839   & 2,884  \\ \hline
	\end{tabular}
	\caption{Comparison of metrics different adaptivity modes for the biconvex aerofoil problem.
		}
	\label{tab:biconvex-comparison}
\end{table}

\subsection{Rotating Square Projectile} \label{sec:rotsq}
This problem involves a square projectile of side length \(l_s = \qty{0.02}{\meter}\) rotating about its center in a Mach 6 flow with freestream temperature \(T_\infty = \qty{710}{\kelvin}\), pressure \(p_\infty = \qty{760}{\pascal}\), and velocity \(u_\infty = \qty{1005.0}{\meter\per\second}\). This problem is borrowed from the example problems of Eilmer~\citep{gibbonsEilmerOpensourceMultiphysics2023} repository. The side of the square projectile is \(l_s = \qty{0.02}{\meter}\). The corners of the square are rounded with a radius of \(\SI{1e-3}{\meter}\). The simulation is run till \(t_f = \qty{2.0}{\milli\second}\). The angular velocity of rotation of the projectile is given as,
\begin{equation}
	\omega(t) = A \cos{\frac{2\pi t}{t_f}},
\end{equation}
where \(A = \qty{2000}{\radian\per\second}\).

We consider two cases: one with VSLA-SAS on and the other with VSLA-SAS off. 
The nominal spacing of the incoming particles is \((m/\rho)^{1/d} = \qty{0.0005}{\meter}\) for the case with VSLA-SAS off and \((m/\rho)^{1/d} = \qty{0.00175}{\meter}\) for the case with VSLA-SAS on. The case with VSLA-SAS on is run with \(\Delta s_{\max} = 3.5 \Delta s_{\min} = \qty{5e-3}{\meter}\). The artificial conductivity parameter, $\alpha_u$ is set as 0.5 for this problem. 

The bow shock formed upstream of the projectile and the particle distribution around the projectile for both the cases can be seen in \cref{fig:rotsq}.  When VSLA-SAS is off, the wake region exhibits serious resolution deficiency.
There are large regions devoid of particles. While the density is very low, any approximation in this region will be inaccurate owing to the scarcity of particles.
When VSLA-SAS is on, the particles around the projectile are refined as part of the local adaptivity feature. 
The solution adaptivity feature is demonstrated by the refinement of the particles in the bow shock region formed upstream of the projectile. This convincingly demonstrates the substantial improvements achieved by the proposed adaptivity procedure.
From \cref{fig:rotsq-force} we can see that the forces on the projectile with VSLA-SAS on is in good agreement with the forces on the projectile with VSLA-SAS off and also with Eilmer~\citep{gibbonsEilmerOpensourceMultiphysics2023} results. As shown in \cref{tab:rotsq-time}, the simulation with VSLA-SAS on is able to achieve a speed-up of more than 10 times compared to the case with VSLA-SAS off. The number of particles in the case with VSLA-SAS on is also significantly lower than the case with VSLA-SAS off. This demonstrates the potential of the proposed adaptive resolution method in expediting the simulations.

\IfFileExists{figs_cache_RotatingSquare_p.pdf}{
	\begin{figure}
		\centering
		\includegraphics[width=0.95\textwidth]{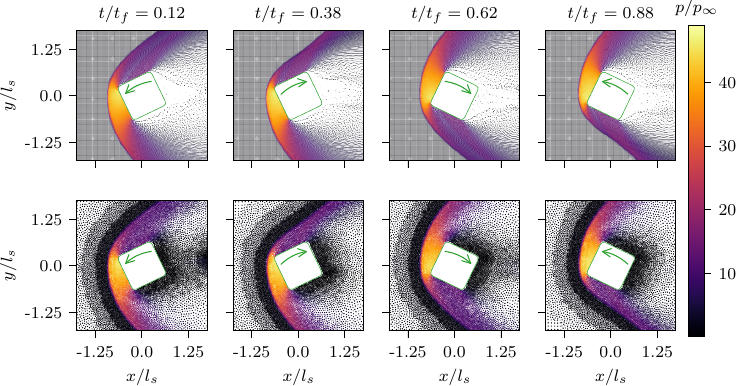}
		\caption{Time snapshots showing particles distribution around rotating square projectile. Top row shows the results without VSLA-SAS and the bottom row shows the results with VSLA-SAS. The arrows denote the sense of rotation of the projectile at the corresponding instant of time.}
		\label{fig:rotsq}
	\end{figure}
}{\textcolor{red}{Figure not found}}

\IfFileExists{figs_cache_RotatingSquare_force.pdf}{
	\begin{figure}[H]
		\centering
		\includegraphics[width=0.45\textwidth]{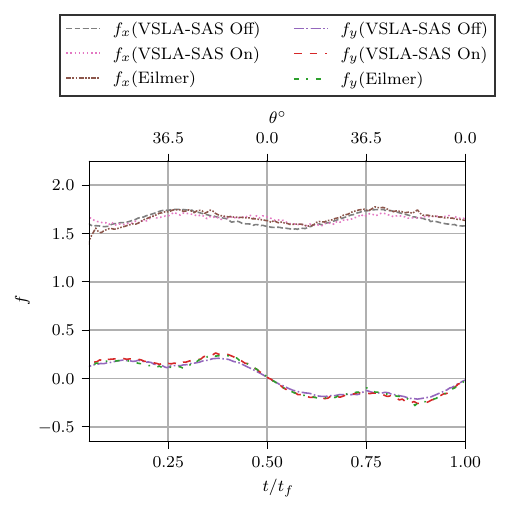}
		\caption{Variation of forces on the rotating square projectile with time. The forces are normalized as \(f = \hat{f}/ (0.5 \rho u^2 l_s)\) where \(\hat{f}\) is the force per unit length. The $f$ here is not to be confused with that in \cref{eq:adami-shepard}. The subscripts \(x\) and \(y\) denote the x and y components of the force, respectively.}
		\label{fig:rotsq-force}
	\end{figure}
}{\textcolor{red}{Figure not found}}

\begin{table}[H]
	\centering
	\begin{tabular}
		{|c|>{\centering\arraybackslash}p{0.15\textwidth}|>{\centering\arraybackslash}p{0.15\textwidth}|c|}
		\hline
		\multirow{2}{*}{Metric} & \multicolumn{2}{c|}{VSLA-SAS} & \multirow{2}{*}{Change}                     \\ \cline{2-3}
		                        & Off                              & On                     &                   \\ \hline \hline
		Final particles         & 139821                        & 22683                    & 83.78\% reduction  \\ \hline
		Time steps               & 16444                            & 7225                     & 56.06\% reduction  \\ \hline
		Simulation time (s)     & 17724                          & 1675                  & 10.58 times faster \\ \hline
	\end{tabular}
	\caption{Comparison of metrics with and without VSLA-SAS for the rotating square projectile problem.}
	\label{tab:rotsq-time}
\end{table}

\subsection{Apollo Reentry Capsule} \label{sec:apollo}

\begin{figure}[H]
    \centering
    \includegraphics[width=0.95\textwidth, trim=0 125 50 350, clip]{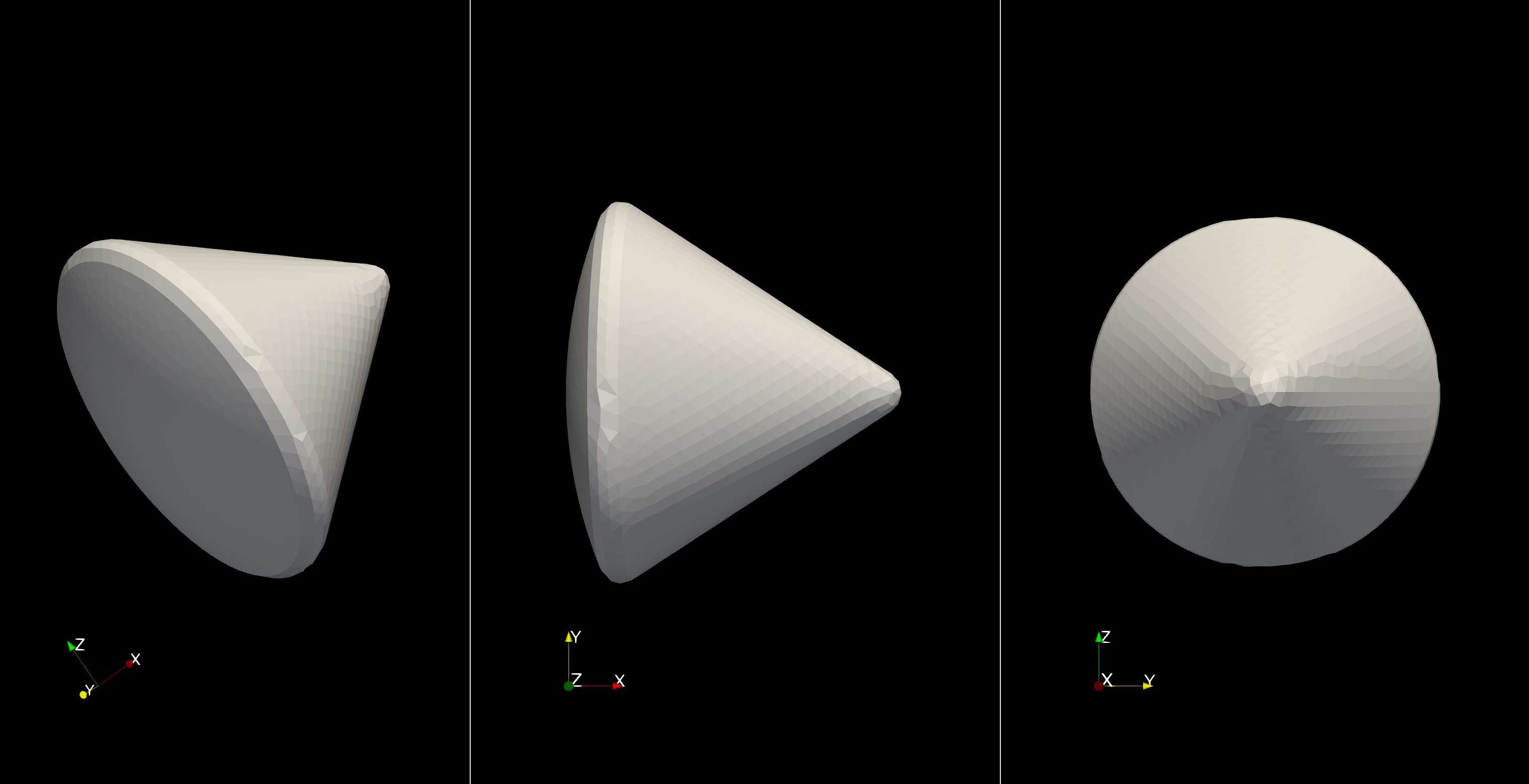}
    \caption{Representation of the Apollo reentry capsule geometry.}
    \label{fig:apollo-geom}
\end{figure}

In this problem, Mach 2.5 flow over a simplified 3D Apollo reentry capsule geometry~\citep{mossDSMCSimulationsApollo2006} is considered. 
This problem aims to demonstrate that the proposed adaptivity procedure can be used for real-world 3D problems. The geometry is shown in \cref{fig:apollo-geom}.
The details of the setup employed for this are available in our previous work~\citep{villodiRobustSolidBoundary2024}.

We showcase two results: one without and the other with adaptive refinement.
For the simulation without adaptive refinement, we use a nominal particle spacing, \((m/\rho)^{1/d} = \qty{0.01}{\meter}\).
As mentioned in our previous work~\citep{villodiRobustSolidBoundary2024}, the particle spacing could not have been much smaller than the chosen spacing as it would have led to a prohibitively huge simulation time, given our computational capabilities.
This being the case, we intended to use the adaptive refinement procedure to decrease the number of particles in the domain by restricting refinement to regions around the body and shocks.
Doubling the nominal spacing at the inlet means that the particle being introduced at the inlet for the adaptively resolved case has 8 times the volume and mass compared to its counterpart from the non-adaptively resolved case.
This would mean a ratio of \(\Delta s_{\max} / \Delta s_{min} = 2\) for the adaptively resolved case.
Thus, the associated volume of the refined region is expected to match the volume of the particles being introduced at the inlet for the non-adaptively refined case.

The results are shown in \cref{fig:a3d-mach}.
In this figure, we have plotted the points such that the size of the points is representative of the volume of the particle.
So, it can clearly be seen that the incoming particles are coarse, and the particles around the body are refined.
The coefficient of force tangential to the flow is also more accurately predicted in the adaptively resolved case as shown in \cref{tab:a3d-cd}.

\newcommand{\cdadapt}{1.5267038151065417}
\newcommand{\cdnoadapt}{1.5426425983828058}
\newcommand{\cdeilmer}{1.4228605452919434}
\begin{table}[H]
	\centering
	\pgfkeys{/pgf/number format/.cd,fixed,fixed zerofill,precision=3}
	\begin{tabular}{|c|c|}
		\hline
		\multicolumn{1}{|c|}{Case}  & \multicolumn{1}{c|}{\(F_x\)}    \\ \hline \hline
		Without Adaptive Resolution & \pgfmathprintnumber{1.5426425983828058} \\ \hline
		With Adaptive Resolution    & \pgfmathprintnumber{1.5267038151065417}   \\ \hline
		Eilmer~\citep[from ][]{villodiRobustSolidBoundary2024}                      & \pgfmathprintnumber{1.4228605452919434}  \\ \hline
	\end{tabular}
	\caption{Non-dimensional force on the Apollo reentry capsule in the streamwise direction. \(F_x = \hat{F}_x / (0.5 \rho u^2 A)\) where \(\hat{F}_x\) is the force in the streamwise direction, \(A = \pi l_s^2\) and \(l_s = \qty{1.99}{\meter}\).}
	\label{tab:a3d-cd}
\end{table}

\IfFileExists{figs_cache_Apollo3D_mach.pdf}{
	\begin{figure}[H]
		\centering
		\includegraphics[width=0.9\textwidth]{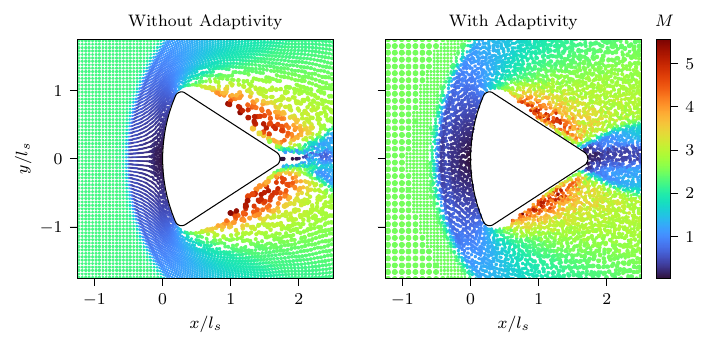}
		\caption{Apollo reentry capsule results showing particles coloured by Mach number.
			The size of the scattered points is representative of the volume of the particle. The plotted points are from a slice of thickness 0.30 m about the \(z=0\) plane. 
			The result on the left is without adaptive resolution, and the result on the right is with adaptive resolution.
		}
		\label{fig:a3d-mach}
	\end{figure}
}{\textcolor{red}{Figure not found}}

\begin{table}[H]
	\centering
	\begin{tabular}
		{|c|>{\centering\arraybackslash}p{0.15\textwidth}|>{\centering\arraybackslash}p{0.15\textwidth}|c|}
		\hline
		\multirow{2}{*}{Metric} & \multicolumn{2}{c|}{VSLA-SAS} & \multirow{2}{*}{Change}                     \\ \cline{2-3}
		                        & Off                              & On                     &                   \\ \hline \hline
		Final particles         & 1725933                          & 678074                   & 60.71\% reduction  \\ \hline
		Time steps               & 1554                            & 1218                     & 21.62\% decrease  \\ \hline
		Simulation time (s)     & 133186                          & 59164                  & 2.25 times faster \\ \hline
	\end{tabular}
	\caption{Comparison of metrics with and without VSLA-SAS for the Apollo reentry problem.}
	\label{tab:apollo-comparison}
\end{table}

\section{Summary and concluding remarks} \label{sec:conclusions}
The presented work focuses on the development of an adaptive resolution procedure for improving the accuracy and efficiency of compressible flow simulations with \gls{sph}.
The proposed adaptivity procedure is capable of local adaptivity and solution adaptivity.
The local adaptivity feature allows for the refinement of particles around bodies or other predefined regions.
The solution adaptivity feature allows for the refinement of particles in interesting regions based on the instantaneous solution.

The proposed adaptive resolution procedure is demonstrated on various problems, including the oscillating piston chamber, double Mach reflection, shock diffraction, compression corner, Noh's implosion, hypersonic cylinder, and biconvex airfoil.
The results validate the effectiveness of the adaptive resolution procedure, showing significant improvements in solution quality.
While the adaptivity algorithm adds some computational overhead, reducing the number of particles effectively makes the simulations faster and less resource-demanding. For example, using the basic adaptivity procedure without local or solution adaptivity, the number of particles in the shock diffraction problem is reduced by 25.7\%, and a speed-up of 1.87 times was realized. In the biconvex airfoil problem, with local and solution adaptivity, the number of particles is reduced by 72.7\%, and a speed-up of 2.05 times was observed compared to the basic adaptivity procedure. 
Similarly, in the Apollo reentry capsule problem becomes faster by 2.25 times with VSLA-SAS compared no adaptivity. The number of particles is reduced by 60.71\%. In the rotating square projectile problem, the adaptive resolution procedure delivered a speedup 10.58 times compared to the non-adaptive simulation. The number of particles were reduced by 83.78\%. This demonstrates the potential of the adaptive resolution procedure to significantly accelerate computations while minimizing the use of computational resources.

The choice of the inclusion of the various features like shock aware shifting or solution adaptivity is problem-dependent. For the double Mach reflection problem, simple volume adaptivity suffices. As we encounter larger density variations in the oscillating piston chamber problem, the particle shifting becomes necessary. With large density variations and the presence of shocks in the shock diffraction problem, shock-aware particle shifting becomes necessary. The rest of the problems demonstrate further improvements and savings with the local and solution adaptivity. In the oscillating piston chamber problem, the pressure obtained from the simulation closely matches the analytical solution, and the variation in the number of particles in the chamber matches the results from the literature.
In the double Mach reflection problem, the adaptivity procedure resolves major artefacts like clumping and kinked Mach stem, allowing for finer resolution without issues.
In the shock diffraction problem, the adaptive resolution effectively resolves issues like particle clumping and resolution deficiency, demonstrating the necessity of adaptive resolution in low-density regions.
The compression corner problem illustrates the solution adaptivity mechanism, showing the ability to capture sharper shocks.
Noh's implosion problem demonstrated improved post-shock density and sharper shock using solution adaptivity on a moving shock. This is a particularly hard problem with infinite Mach number, where we have shown the introduction of some noise.  Otherwise, in terms of quality of results, the adaptive resolution is a faster way to obtain similar or better results, i.e., without sacrificing accuracy. 
The hypersonic cylinder and biconvex airfoil problems demonstrate the local adaptivity and solution adaptivity features, with the results showing good agreement with theoretical solution and literature data.

Conservation is highly valued in compressible \gls{cfd} for accurate shock speeds. 
The proposed splitting and merging procedures conserve mass, volume, and momentum and thermal energy. 
Splitting can be made kinetic energy and angular momentum conservative by turning off the extrapolation. 
The same cannot be said for merging.
However, loss of conservation can be minimized during merging by following \cite{havasi-tothParticleCoalescingAngular2020,sunConservativeParticleSplitting2023}. 
Our test cases show that the loss of conservation due to any of the introduced techniques is not significant and barely affects the solution quality. Therefore, the argument that the reduction in computational requirements and speed-up obtained by the adaptive resolution procedure significantly outweighs the potential downsides due to loss of conservation remains valid.

Overall, the proposed adaptive resolution procedure demonstrates the capability to improve solution quality and reduce computational costs.
A formal convergence study of the scheme and proposed adaptive resolution procedure represents a promising direction for future work.
Further research could also explore the integration of ideas from \citet{garcia-senzConservativeDensitybasedSmoothed2022, havasi-tothParticleCoalescingAngular2020, sunConservativeParticleSplitting2023} for improving the conservation of the basic discretization scheme and the adaptive resolution procedure.
Noise reduction in the solution is another aspect that could be explored in the future.

\section{Acknowledgements}
The authors acknowledge the use of the computing resources of the ACE Facility, Department of Aerospace Engineering, IIT Bombay. 
The authors also acknowledge National Supercomputing Mission (NSM) for providing computing resources of `PARAM RUDRA' at IIT Bombay which is implemented by C-DAC and supported by the Ministry of Electronics and Information Technology (MeitY) and Department of Science and Technology (DST), Government of India.
\bibliographystyle{model6-num-names}
\bibliography{references}

\end{document}